\documentclass[preprint]{vgtc}                          

\ifpdf                  
  \pdfoutput=1\relax                 
  \usepackage{graphicx}
  \DeclareGraphicsExtensions{.pdf,.png,.jpg,.jpeg}
\else
  \usepackage{graphicx}
  \DeclareGraphicsExtensions{.eps}
\fi

\graphicspath{{figures/}{pictures/}{images/}{./}} 

\usepackage{microtype}
\PassOptionsToPackage{warn}{textcomp}
\usepackage{textcomp}
\usepackage{mathptmx}
\usepackage{times}

\usepackage{cite}                      
\usepackage{tabu}               
\usepackage{booktabs}

\usepackage{paralist}
\usepackage[font={small,sf},caption=false,captionskip=1pt]{subfig}
\usepackage{color, soul}
\usepackage[font={scriptsize,sf},tableposition=top]{caption}
\usepackage{float}
\floatstyle{plaintop}
\restylefloat{table}
\usepackage{siunitx}
\usepackage{colortbl}
\usepackage{sparklines}

\usepackage{multicol}
\usepackage[bookmarks=false]{hyperref}

\preprinttext{To appear in IEEE Conference on Visual Analytics Science and Technology (VAST) 2020}

\onlineid{1135}
\vgtccategory{Research}

\vgtcinsertpkg

\newcommand{\Title}{
A Visual Analytics Framework for Contrastive Network Analysis
}
\title{\Title}

\author{
Takanori Fujiwara\thanks{e-mail: tfujiwara@ucdavis.edu},\textsuperscript{1}\\ %
\and 
Jian Zhao\thanks{e-mail: jianzhao@uwaterloo.ca},\textsuperscript{2}\\ %
\and 
Francine Chen\thanks{e-mail: francine@acm.org}\\ %
\and
Kwan-Liu Ma\thanks{e-mail: klma@ucdavis.edu},\textsuperscript{1}\\ %
}
\affiliation{ 
\vspace{-10pt}
\scriptsize \text{\textsuperscript{1}University of California, Davis}, \scriptsize \text{\textsuperscript{2}University of Waterloo}
}

\shortauthortitle{Fujiwara \MakeLowercase{\textit{et al.}}: \Title}

\abstract{
A common network analysis task is comparison of two networks to identify unique characteristics in one network with respect to the other. For example, when comparing protein interaction networks derived from normal and cancer tissues, one essential task is to discover protein-protein interactions unique to cancer tissues. However, this task is challenging when the networks contain complex structural (and semantic) relations. To address this problem, we design ContraNA, a visual analytics framework leveraging both the power of machine learning for uncovering unique characteristics in networks and also the effectiveness of visualization for understanding such uniqueness. The basis of ContraNA is cNRL, which integrates two machine learning schemes, network representation learning (NRL) and contrastive learning (CL), to generate a low-dimensional embedding that reveals the uniqueness of one network when compared to another. ContraNA provides an interactive visualization interface to help analyze the uniqueness by relating embedding results and network structures as well as explaining the learned features by cNRL. We demonstrate the usefulness of ContraNA with two case studies using real-world datasets. We also evaluate ContraNA through a controlled user study with 12 participants on network comparison tasks. The results show that participants were able to both effectively identify unique characteristics from complex networks and interpret the results obtained from cNRL.
}

\keywords{Contrastive learning, network representation learning, interpretability, network comparison, visual analytics.}

\nocopyrightspace

\newcommand{\name}{ContraNA}

\newcommand{\etal}{et~al.}
\newcommand{\q}[1]{\textit{``#1''}}
\newcommand{\p}[1]{\textit{p#1}}
\newcommand{\f}[1]{\textsf{\scriptsize #1}}

\begin{document}
\firstsection{Introduction}
\maketitle

\newcommand{\Graph}[1]{G_#1}
\newcommand{\GraphT}{\Graph{T}}
\newcommand{\GraphB}{\Graph{B}}
\newcommand{\Adj}[1]{\mathbf{A}_#1}
\newcommand{\AdjT}{\Adj{T}}
\newcommand{\AdjB}{\Adj{B}}
\newcommand{\Attr}[1]{\mathbf{P}_#1}
\newcommand{\AttrT}{\Attr{T}}
\newcommand{\AttrB}{\Attr{B}}
\newcommand{\nNodes}[1]{n_#1}
\newcommand{\nNodesT}{\nNodes{T}}
\newcommand{\nNodesB}{\nNodes{B}}
\newcommand{\nAttrs}[1]{m_#1}
\newcommand{\nAttrsT}{\nAttrs{T}}
\newcommand{\nAttrsB}{\nAttrs{B}}
\newcommand{\nEdges}[1]{l_#1}
\newcommand{\nEdgesT}{\nEdges{T}}
\newcommand{\nEdgesB}{\nEdges{B}}
\newcommand{\nNRLFeats}{d}
\newcommand{\nCNRLFeats}{{d'}}
\newcommand{\FeatMat}[1]{\mathbf{X}_#1}
\newcommand{\FeatMatT}{\FeatMat{T}}
\newcommand{\FeatMatB}{\FeatMat{B}}
\newcommand{\ProjMat}{\mathbf{W}}
\newcommand{\ContRepr}[1]{\mathbf{Y}_#1}
\newcommand{\ContReprT}{\ContRepr{T}}
\newcommand{\ContReprB}{\ContRepr{B}}

\newcommand{\BaseFeat}{{\rm \bf x}}
\newcommand{\RelFunc}{f}
\newcommand{\RelFeatOpe}[2]{\Phi^{#1}_{#2}}
\newcommand{\DepthRelFunc}{h}
\newcommand{\SummaryMeasure}{S}
\newcommand{\Mean}{\rm mean}
\newcommand{\Sum}{\rm sum}
\newcommand{\Max}{\rm max}
\newcommand{\Lpnorm}{L^2{\rm norm}}
\newcommand{\Cov}[1]{\mathbf{C}_#1}
\newcommand{\CovT}{\Cov{T}}
\newcommand{\CovB}{\Cov{B}}
\newcommand{\ContParam}{\alpha}
\newcommand{\ContConst}{\epsilon}

\newcommand{\DolphinNwID}{N1}
\newcommand{\KarateNwID}{N2}
\newcommand{\RandomNwID}{N3}
\newcommand{\PriceNwID}{N4}
\newcommand{\PtoPNwID}{N5}
\newcommand{\PriceTwoNwID}{N6}
\newcommand{\EPriceNwID}{N7}
\newcommand{\CombinedAPMSNwID}{N8}
\newcommand{\LCMultiNwID}{N9}
\newcommand{\SchFirstDayNwID}{N10}
\newcommand{\SchSecondDayNwID}{N11}

\newcommand{\DolphinVsKarate}{F1}
\newcommand{\PriceVsRandom}{F2}
\newcommand{\RandomVsPrice}{F3}
\newcommand{\PtoPVsPriceTwo}{F4}
\newcommand{\PtoPVsEPrice}{F5}
\newcommand{\LCMultiVsCombinedAPMS}{F6}
\newcommand{\SchSecondVsFirst}{F7}

\newcommand{\SE}{\mathrm{SE}}

A network is a common form for modeling various types of relationships in real-world applications, such as social connections~\cite{bezerianos2010graphdice,crnovrsanin2014visualization}, biological interactions~\cite{chen2004content,gehlenborg2010visualization}, and supercomputer communications~\cite{bhanot2005optimizing}. 
In practice, comparative analysis of two networks is vital~\cite{emmert2016fifty,tantardini2019comparing}, especially for the identification of the uniqueness of one network compared to another.
We call this task \emph{contrastive network analysis}. 
For example, when studying the effect of Alzheimer's disease on a human brain~\cite{gaiteri2016genetic}, neuroscientists want to find unique functional connections in the brain network of a patient with Alzheimer's disease by comparing to that of a healthy subject. 
Also, for researcher collaborations in different disciplines~\cite{lariviere2006canadian}, analysts in a funding agency may want to reveal unique ways of collaboration in the disciplines for decision making.

Despite the demands for network comparison, there is little adequate visual analytics support. 
Most of the existing methods (e.g., \cite{alper2013weighted,shi2015brainquest,kerpedjiev2018higlass}) presuppose the existence of node-correspondence (i.e., pairwise correspondence between nodes in two different networks)~\cite{tantardini2019comparing}. 
This is a critical limitation since we usually do not know such information in advance when the networks are collected from different resources.
One potential solution is identifying the node-correspondence by using network alignment (or graph matching)~\cite{emmert2016fifty,tantardini2019comparing}. 
However, these algorithms notoriously have high computational costs~\cite{emmert2016fifty,tantardini2019comparing}, and thus are only suitable for treating small networks (e.g., 100 nodes).
Also, there may not exist a clear correspondence between nodes. 

Another approach for visual comparison of networks is based on statistical measures (e.g., network density)~\cite{freire2010manynets}, centralities (e.g., degree centrality)~\cite{yoghourdjian2018graph}, graphlets~\cite{kwon2017would}, or a combination of these~\cite{von2009visual}.
For example, with graphlets~\cite{prvzulj2007biological} (small, connected, and non-isomorphic subgraph patterns in a network), the similarities of two networks can be measured by comparing the frequency of appearance of each graphlet in each network~\cite{kwon2017would}. 
While these approaches can provide a (dis)similarity between different networks, they compare networks only based on simple measures, which are often insufficient.
Also, they only provide network-level similarities, and thus cannot compare networks at more detailed levels (e.g., a node-level). 
Without a detailed-level comparison, it is difficult to find which part of a network relates to its uniqueness. 

To address the above problems, we introduce a novel visual analytics framework, \textit{\name{}}, for comparative network analysis, which integrates \textit{contrastive network representation learning (cNRL)}~\cite{icnrl} into interactive visualization.
Empowered by cNRL, our framework allows for discovering unique characteristics of one network by contrasting with another in a comprehensive (i.e., using multiple advanced measures) and detailed (i.e., analyzing a node or subnetwork level) manner without node-correspondence information. 
Specifically, we employ an interpretable version of cNRL (i-cNRL)~\cite{icnrl} to provide human-understandable explanations of discovered characteristics that are further revealed by novel visual representations.
We enhance i-cNRL by designing an interactive visual interface that allows analysts to integrate their domain knowledge into the automated analysis. 
Particularly, we introduce a method to visually identify the uniqueness in one network based on the i-cNRL result, a visual summary to intuitively inform network features that highly contribute to the result, and interactive linkings with the existing network visualizations to explain and refine the result. 

In summary, our main contributions include:
\begin{compactitem}
    \item A cNRL-based visual analytics framework, \name{}, which aims to support a new network analysis approach, named contrastive network analysis, to effectively reveal unique characteristics in one network relative to another.
    \item Enhancements of i-cNRL with a visual interface that provides four major abilities---\textbf{DIIF}: (1) \textbf{D}iscovery of uniqueness in networks, (2) \textbf{I}nterpretability of features generated by i-cNRL, (3) \textbf{I}ntuitive analysis with common visualizations, and (4) \textbf{F}lexibility of adjusting i-cNRL based on analysts' interests. 
    \item Two case studies and a controlled user study with multiple real-world datasets, which assess the effectiveness and usefulness of \name{} for contrastive network analysis.
\end{compactitem}

\section{Background and Related Work}

In this section, we first describe network representation learning and contrastive learning---two machine learning schemes used in \name.
Then, we review the relevant works for visual network comparison.

\subsection{Network Representation and Contrastive Learning}
\label{sec:nrl_cl}

\textit{Network representation learning (NRL)}~\cite{cai2018comprehensive,zhang2018network}, also known as graph embedding, aims to learn low-dimensional latent vectors that represent a network while maximally preserving certain network information, such as the structural and semantic characteristics. 
Once a low-dimensional representation obtained, we can easily and efficiently conduct network analysis tasks (e.g., node classification and link prediction). 
Typical NRL methods include node2vec~\cite{grover2016node2vec}, DeepGL~\cite{rossi2018deep}, and some other deep neural networks~\cite{zhang2018deep}. 
More comprehensive descriptions of NRL methods have been included in several recent surveys~\cite{cai2018comprehensive,zhang2018network}.

\textit{Contrastive learning (CL)}~\cite{zou2013contrastive} focuses on finding patterns that are more salient in one dataset relative to another~\cite{abid2018exploring}.
This is unlike discriminant analysis (e.g., linear discriminant analysis~\cite{JiaNZ09}), which aims to discriminate data points based on their classes. 
Several CL methods have been developed in the machine learning community, such as contrastive versions of latent Dirichlet allocation~\cite{zou2013contrastive}, hidden Markov models~\cite{zou2013contrastive}, and regressions~\cite{ge2016rich}.
CL methods for representation learning have been also introduced~\cite{ge2016rich,abid2018exploring,dirie2019contrastive,abid2019contrastive,severson2019unsupervised}, such as contrastive PCA (cPCA)~\cite{ge2016rich,abid2018exploring} and contrastive variational autoencoder (cVAE)~\cite{abid2019contrastive,severson2019unsupervised}.

Recently, contrastive network representation learning (cNRL)~\cite{icnrl} integrates the above two machine learning schemes to achieve comparative network analysis.
It uses NRL to generate two sets of latent vectors for two networks and then employs CL to perform comparative analysis based on the vectors.
This approach embeds network nodes into a low-dimensional space that reveals the uniqueness of one network compared to another. 
To offer interpretability, \name{} uses a specific cNRL method (see \autoref{sec:analytics_pipeline}) and further provides novel visual analysis capabilities to enable effective network comparison.

\subsection{Visualization for Network Comparison}

There exist three general approaches in visual comparison: juxtaposition, superposition, and explicit encoding~\cite{gleicher2011visual}.
Through a comprehensive survey, Gleicher~\cite{gleicher2018considerations} provided a  framework of considerations for visual comparison, such as tasks, challenges, strategies, and designs. 
Here we review the relevant works in visual network comparison.

\subsubsection{Static Network Comparison}

Comparing multiple static networks has been a classic problem in visualization research.
Alper \etal~\cite{alper2013weighted} presented several superposition designs for node-link and adjacency matrix visualizations to support weighted network comparison.
TileMatrix~\cite{liu2015effects} uses juxtaposition to place the triangular adjacency matrices of two networks onto upper and lower areas of a square matrix.
On the other hand, John \etal~\cite{john2019visual} juxtaposed each pair of weighted links in a matrix cell.
MatrixWave further extended this approach to support the comparison of multi-layer networks~\cite{Zhao15}

Researchers have focused on developing techniques for comparing brain networks due to its special characteristics (e.g., very dense) and importance. 
Shi \etal~\cite{shi2015brainquest} opted to visualize links that are significantly different between two brain networks. 
Yang \etal~\cite{yang2017blockwise} used a clustering algorithm with NodeTrix~\cite{henry2007nodetrix}, a hybrid of node-link and adjacency matrix representations.
Fujiwara \etal~\cite{fujiwara2017visual} enabled the comparison of a larger number of brain networks by providing an overview with dimensionality reduction.
Some other domains have been addressed as well, such as genome interaction~\cite{kerpedjiev2018higlass} and egocentric networks~\cite{liu2017egocomp}.

All the above methods require the information of exact node-correspondence~\cite{emmert2016fifty}, unlike \name{}.
While a few works~\cite{andrews2009visual,koop2013visual,liu2014topicpanorama} applied network alignment~\cite{emmert2016fifty} to find node-correspondence before visualization, they do not scale well due to the computation cost. 

\subsubsection{Dynamic Network Comparison}

Dynamic networks contain nodes and/or links changing over time. 
A comprehensive survey is provided by Beck \etal~\cite{beck2017taxonomy}.  
Here, we focus on the comparison of networks at different timestamps.

One approach is based on the juxtaposition of networks at different timestamps.
Federico \etal~\cite{federico2011visual} applied a 2D network layout that produces stable node positions across time and then juxtaposed networks at multiple time points in a 2.5D view.
On the other hand, TimeArcs~\cite{dang2016timearcs} lays out a network at each time point in 1D, and uses an arc diagram to display links. 
A wall-size display was used to juxtapose an array of networks~\cite{lee2019dynamic}.
Moreover, animated transitions have been employed, which can be viewed as juxtaposition in the temporal domain, e.g., GraphDiaries~\cite{bach2013graphdiaries} and DiffAni~\cite{rufiange2013diffani}.

Moreover, several works summarize a dynamic network based on the similarity of the network at each timestamp. 
For example, Small MultiPiles~\cite{bach2015small} groups similar weighted adjacency matrices across consecutive time points and then shows a representative matrix for each group.
EgoLines~\cite{zhao2016egocentric} effectively visualizes a $k$-hop dynamic egocentric network with a ``subway map'' metaphor.
van den Elzen \etal~\cite{van2016reducing} utilized dimensionality reduction to overview the similarities of networks across time.
A similar approach was used to visualize dynamic brain networks~\cite{bach2016time} and compare dominance variation in animal groups~\cite{coelho2020peckvis}.

Lately, researchers have started to utilize time-series or topological analysis to summarize or identify important trends in a dynamic network. 
Examples include using graph wavelet transform to classify nodes~\cite{dal2017wavelet} and persistent homology to capture topological changes~\cite{hajij2018visual}.
Fujiwara \etal~\cite{fujiwara2018visual} applied change point detection~\cite{aminikhanghahi2017survey} to segment a dynamic network and generate summaries.
Several works extended this approach in other cases, such as visualizing streaming networks~\cite{kesavan2020visual,murugesan2019state}.

Again, the above methods still require the information of node-correspondence. To overcome this limitation, we utilize NRL to capture the network's topological and semantic features.

\subsubsection{Comparison without Node-Correspondence}

Several systems were developed to support network comparison without the limitation of knowing node-correspondence.
ManyNets~\cite{freire2010manynets} uses a tabular interface to list several basic network statistics (e.g., degree centrality) for each network.
von Landesberger \etal~\cite{von2009visual} used graphlet frequencies and other network-statistics measures and to generate a self-organization map for arranging networks on a 2D grid.  
A similar approach was used by Harrigan \etal~\cite{harrigan2012egonav} to visualize egocentric networks, and by Kwon \etal~\cite{kwon2017would} to show similar networks given an input network.
In addition to node-level features, Gove~\cite{gove2019gragnostics} suggested network-level features (e.g., density) that are easier to interpret and faster to compute.
Along this line, Graph Thumbnails~\cite{yoghourdjian2018graph} uses the $k$-core number in a nested circle packing representation of networks.

While the above methods can be used for comparing networks without node-correspondence, they lack the ability to compare networks from multiple levels. \name{} addresses this by employing the state-of-the-art NRL method, allowing for comparison at both node and subgraph levels. 
Further, by leveraging CL, we focus on revealing the uniqueness in one network relative to another, which is different from the purpose of the above works (i.e., identifying similarities of networks). 
To the best of our knowledge, the only work using CL for visual analytics~\cite{fujiwara2020supporting} focuses on high-dimensional data. However, \name{} focuses on comparative analysis of networks.

In sum, the existing methods have limited flexibility in use due to the requirement of node-correspondence or to insufficient analysis ability due to the absence of multiple level comparison.
\name{} addresses these issues by utilizing cNRL, which we describe in \autoref{sec:analytics_pipeline}. 
Then, with interactive visualizations, \name{} further supplements cNRL's limitations that are identified in \autoref{sec:dc}.

\section{Contrastive Network Representation Learning}
\label{sec:analytics_pipeline}

Here, we provide a brief introduction to the core analysis method used in \name{}: contrastive network representation learning (cNRL)~\cite{icnrl}. 

\begin{figure}[tb]
	\centering
	\captionsetup{farskip=0pt} 
    \includegraphics[width=0.975\linewidth]{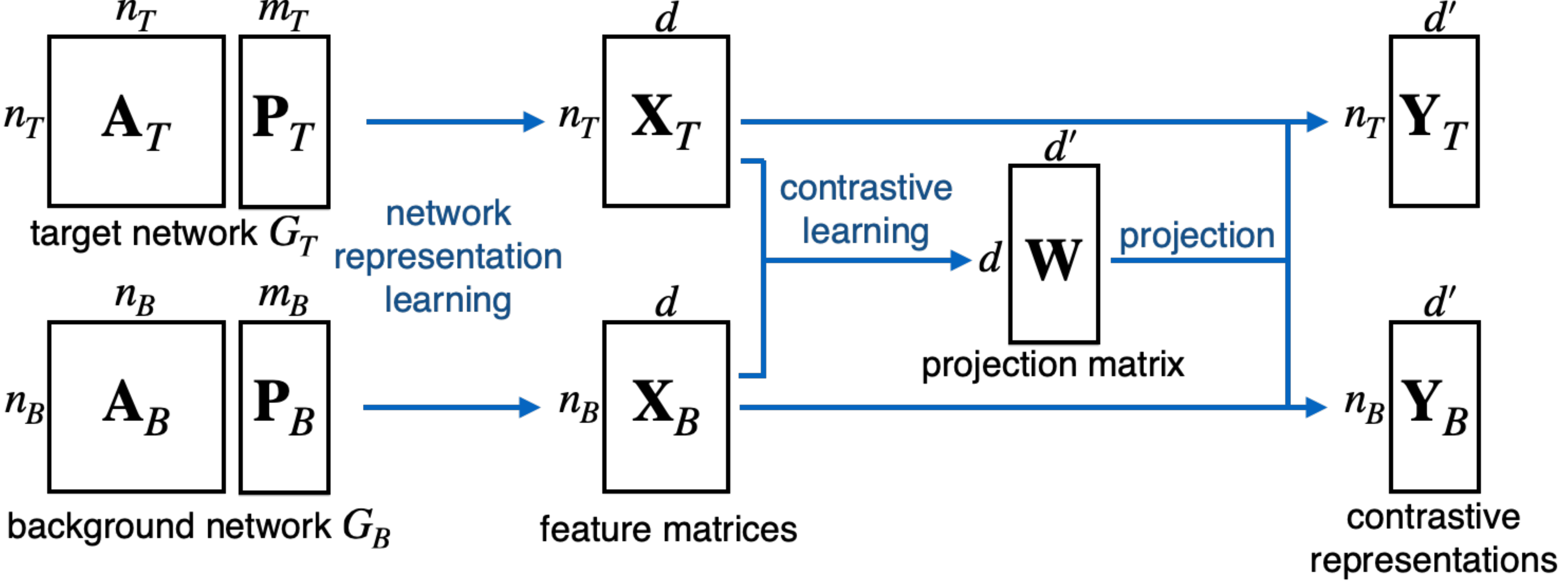}
    \caption{General architecture for cNRL.}
	\label{fig:architecture}
\end{figure}

\subsection{cNRL Architecture}

\autoref{fig:architecture} shows an overview of the cNRL architecture. 
Given two networks, a target network $\GraphT$ and a background network $\GraphB$, the objective of cNRL is learning a contrastive representation $\ContReprT$ that reveals unique characteristics in $\GraphT$ relative to $\GraphB$.
To achieve this, cNRL employs a two-step embedding process: (1) NRL, which obtains network features of $\GraphT$ and $\GraphB$, and (2) CL, which generates a contrastive representation from the network features. 

The input networks $\GraphT$ and $\GraphB$ can be any combination of undirected or directed, unweighted or weighted, and non-attributed or attributed networks. 
In their adjacency matrices $\AdjT$ and $\AdjB$, the numbers of $\GraphT$ and $\GraphB$ nodes, $\nNodesT$ and $\nNodesB$, do not have to be the same. 
Similarly, when $\GraphT$ and $\GraphB$ are attributed, the numbers of attributes $\nAttrsT$ and $\nAttrsB$ in matrices of node attributes $\AttrT$ and $\AttrB$ may be different. 
The first embedding with NRL produces target and background networks' feature matrices $\FeatMatT$ and $\FeatMatB$, where both $\FeatMatT$ and $\FeatMatB$ need to have the same $\nNRLFeats$ features.
Using NRL, cNRL preserves the target and background network information in $\FeatMatT$ and $\FeatMatB$ with explicit and comprehensive network features.
Based on $\FeatMatT$ and $\FeatMatB$, the second embedding using CL generates a projection matrix $\ProjMat$ of $\nNRLFeats$ rows and $\nCNRLFeats$ columns ($\nCNRLFeats \leq \nNRLFeats$) and then $\GraphT$ and $\GraphB$'s contrastive representations $\ContReprT$ and $\ContReprB$ can be produced by multiplying $\FeatMatT$ and $\FeatMatB$ with $\ProjMat$, respectively. 
Through CL, the contrastive representation $\ContReprT$ captures relationships (e.g., the network structural differences among network nodes) that appear in $\GraphT$ but do not appear in $\GraphB$.

The cNRL architecture provides flexibility in selection of methods for both NRL and CL. 
For NRL, we can choose any algorithm that can produce the same features across networks, such as \textit{inductive} NRL methods (e.g., GraphSAGE~\cite{hamilton2017inductive} and DeepGL~\cite{rossi2018deep}), which learn transferable knowledge from a training network to other networks. 
For CL, we can choose any method designed for representation learning. 
See the work about cNRL~\cite{icnrl} for further details about available algorithm options for NRL and CL.

\subsection{Interpretable cNRL (i-cNRL)}
\label{sec:icnrl}

In \name{}, to interactively examine the identified unique characteristics with human-understandable explanations, we specifically employ an interpretable version of cNRL, called \emph{i-cNRL}, where DeepGL~\cite{rossi2018deep} and cPCA~\cite{abid2018exploring} are used as the NRL and CL methods, respectively.
Providing interpretability is the core feature of \name{} as it helps the analysts understand the meaning of unique characteristics found in one network and the reason why such uniqueness can be seen in only that network.

\textbf{NRL with DeepGL.}
Using DeepGL~\cite{rossi2018deep} as NRL, i-cNRL generates feature matrices $\FeatMatT$ and $\FeatMatB$ with interpretable network features. 
The features consist of the \textit{base feature} $\BaseFeat$ and \textit{relational function} $\RelFunc$. 

A base feature $\BaseFeat$ is a measure we can obtain for each node, such as \f{in-, out-degree, degeneracy} (\f{$k$-core numbers})~\cite{newman2018networks}, \f{PageRank}~\cite{newman2018networks}, or an attribute (e.g., gender of a node in a social network).

A relational function $\RelFunc$ is a combination of \textit{relational feature operators} (RFOs), each of which summarizes base feature values of one-hop neighbors of a node. 
For example, the operator can be a computation of the \textit{mean, sum, maximum} base feature values of one-hop neighbors of a node.
Also, the neighbors can be either \textit{in-, out-, total-neighbors}. Together with the summary measure $S$, the operators are denoted $\RelFeatOpe{-}{S}$, $\RelFeatOpe{+}{S}$, and $\RelFeatOpe{}{S}$, respectively. 
For example, $\RelFeatOpe{-}{\Mean}(\BaseFeat)$ computes the mean $\BaseFeat$ of the in-neighbors of a node. 
Moreover, the RFO can be applied repeatedly. 
For example, $\RelFunc = (\RelFeatOpe{+}{\Mean} \circ \RelFeatOpe{-}{\Max})(\BaseFeat)$ first computes the maximum $\BaseFeat$ of in-neighbors for each out-neighbor of a node and then produces the mean of these maximum values. 

During the learning process, from the user-input base features, RFOs, and the maximum number of hops to be considered, DeepGL evaluates combinations of these inputs and parameters and automatically selects important network features for preserving the topological (and semantic) information. 
For example, when using \f{in-degree} and \f{out-degree} as base features, $\RelFeatOpe{}{\Mean}$ and $\RelFeatOpe{}{\Sum}$ as RFOs, and 2 as the maximum number of hops, DeepGL may generate the network features \{\f{in-degree, out-degree,} $\RelFeatOpe{}{\Mean}$(\f{in-degree}),  $\RelFeatOpe{}{\Sum}$(\f{in-degree}), $\RelFeatOpe{}{\Mean} \circ \RelFeatOpe{}{\Mean}$(\f{in-degree})\}. 

\textbf{CL with cPCA.} 
From the target and background feature matrices $\FeatMatT$ and $\FeatMatB$, cPCA~\cite{abid2018exploring} produces \textit{contrastive principal components (cPCs)}, which are analogous to principal components (PCs) in ordinary PCA~\cite{jolliffe1986principal}. 
cPCs are low-dimensional representative directions where $\FeatMatT$ has high variance but $\FeatMatB$ has low variance.
That is, $\ContReprT$, an embedding of $\FeatMatT$ with cPCs, depicts unique characteristics (with the consideration of variance) of a target network $\GraphT$ relative to a background network $\GraphB$.

cPCA requires one hyperparameter $\ContParam$ $(0 \leq \ContParam \leq \infty)$, called a \textit{contrast parameter}. 
The contrast parameter $\ContParam$ controls the trade-off between having high target variance and low background variance in cPCs. 
When $\ContParam = 0$, cPCs only maximize the variance of $\FeatMatT$, the same as those in classical PCA.
As $\ContParam$ increases, cPCs place greater emphasis on directions that reduce the variance of $\FeatMatB$.
Because $\ContParam$ has a strong impact on the result, researchers have developed semi-automatic~\cite{abid2018exploring} and automatic~\cite{icnrl} selection of $\ContParam$. 

Similar to PCs in PCA, cPCs are represented as linear transforms of $\nNRLFeats$ features of $\FeatMatT$ and $\FeatMatB$. 
Analogous to PC loadings, cPCA provides \textit{cPC loadings}~\cite{fujiwara2020supporting}, which indicate how strongly each of the $\nNRLFeats$ input features contributes to the corresponding cPC.
By examining a list of $\nNRLFeats$ learned features via NRL and cPC loadings, we can understand the relationships between the $\nNRLFeats$ features and cPCs; and we can also interpret the contrastive representation $\ContReprT$.

\section{Design Considerations}
\label{sec:dc}

The aforementioned i-cNRL can generate a contrastive representation which highlights the uniqueness of a target network.
However, to thoroughly understand the uniqueness, we opt to empower the automated analysis with interactive visualization, which can tightly integrate the knowledge and adaptability of human experts with the statistical learning of machines~\cite{tam2016analysis,sacha2016visual}.
We comprehensively identify a set of limitations to i-cNRL for contrastive network analysis in depth, which leads to the following design considerations for our visual analytics framework, \name{}.
In general, we aim to amplify the \textbf{D}iscovery, \textbf{I}nterpretability, \textbf{I}ntuitiveness, and \textbf{F}lexibility (\textbf{DIIF}) in visual contrastive network analysis.

\textbf{DC1:} \textit{\textbf{Support the discovery} of whether a target network is unique compared to a background network, and which part of the network relates to the uniqueness.}
The uniqueness of a target dataset relative to the base is embedded in the contrastive representation $\ContReprT$ generated by CL-based representation learning methods, including cNRL.
Many previous works attempted to display this data to reveal the uniqueness~\cite{abid2018exploring,abid2019contrastive,dirie2019contrastive}.
However, because $\ContReprT$ only contains the information of the target network $\GraphT$, reviewing only $\ContReprT$ is not sufficient to understand how well the CL method finds uniqueness. 
Also, it is difficult to identify which data points (i.e., network nodes in our case) highly relate to the found uniqueness. 
The visual analytics framework should support discovering the uniqueness and the associated nodes by presenting the information in both the target and background networks.

\begin{figure*}[tb]
    \centering
    \includegraphics[width=0.97\linewidth]{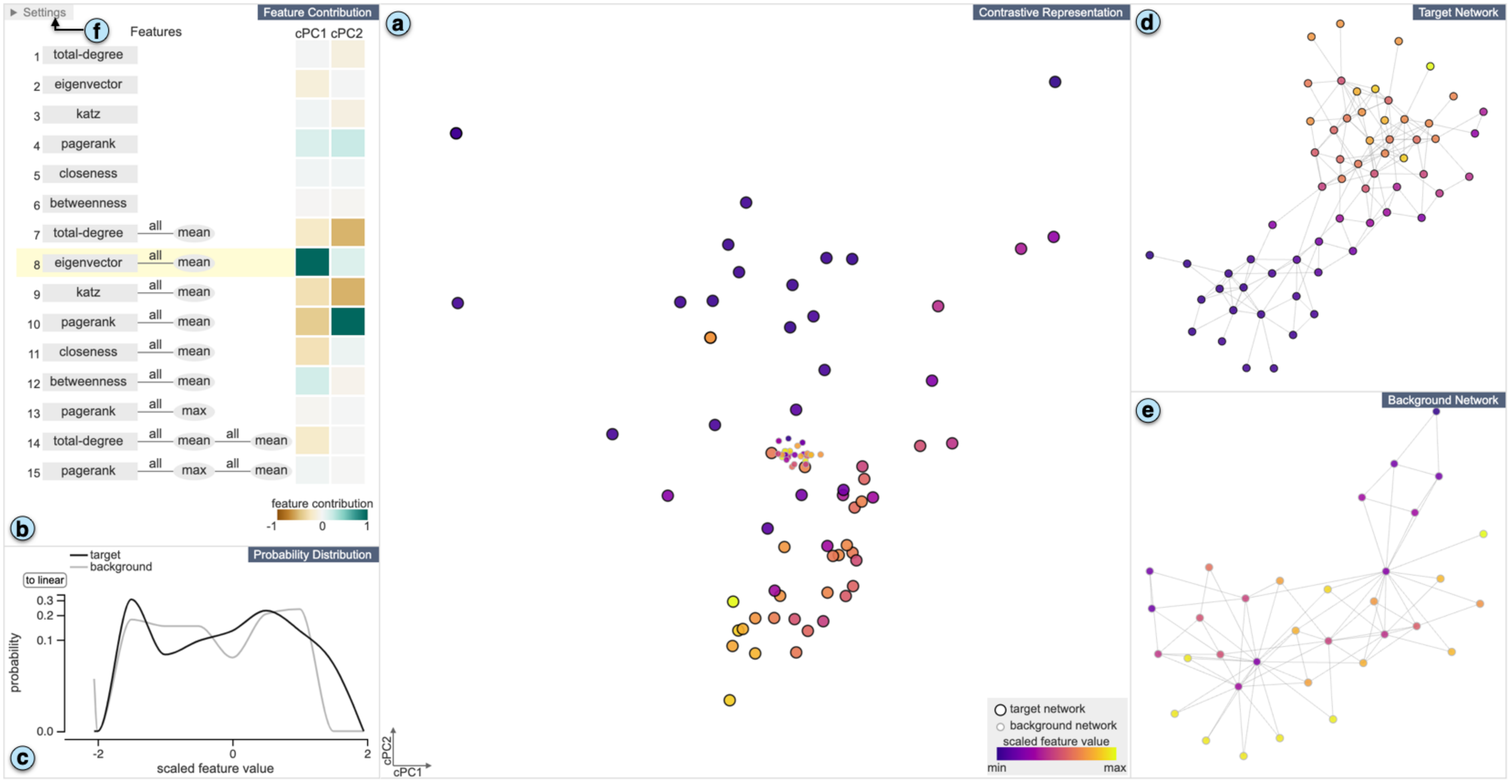}
    \caption{The analyst is using \name{} to conduct a contrastive analysis of the Dolphin social network~\cite{lusseau2003bottlenose} (the target network) and the Zachary’s karate club network~\cite{zachary1977information} (the background network). 
    (a) A contrastive representation view shows contrastive representations of target and background networks. 
    (b) A feature contribution view visualizes network features generated by DeepGL and their contributions to each cPC (i.e., scaled cPC loadings). 
    (c) A probability distribution view depicts target and background networks' probability distributions of the selected network feature in (b).
    (d)(e) A network layout view draws laid-out target and background networks, respectively.
    (f) The analyst can change several settings of the algorithm and visualizations from the drop-down menu.}
    \label{fig:system_overview}
\end{figure*}

\textbf{DC2:} \textit{\textbf{Enhance the interpretability} of the features learned by NRL and the cPCs generated by CL.}
Investigating the relationships among the network features, cPCs, and the representation $\ContReprT$ is important to interpret the uniqueness of $\GraphT$.
While i-cNRL is designed to provide interpretable network features and cPCs, understanding them from i-cNRL's direct outputs is not straightforward. 
For example, DeepGL could generate a sophisticated relational function such as $(\RelFeatOpe{+}{\Sum} \circ \RelFeatOpe{}{\Max} \circ \RelFeatOpe{-}{\Mean})(\BaseFeat)$.
Moreover, examining cPC loadings for each feature would be time-consuming when DeepGL produces many network features. 
The framework should provide visualizations to facilitate easy understanding of the above information.

\textbf{DC3:} \textit{\textbf{Offer intuitiveness} in understanding a target network's uniqueness by relating it to common network visualizations.}
The contrastive representation $\ContReprT$ generated could contain complicated patterns that are difficult to understand. 
Thus, it is not intuitive enough to just view these patterns directly based on the i-cNRL results in the embedding space.
To help analyze such patterns, the framework should provide links between the results of i-cNRL and commonly used visualizations for network analysis, such as laid-out networks and probability distributions of network centralities.

\textbf{DC4:} \textit{\textbf{Provide the flexibility} to interactively adjust the i-cNRL parameters to generate results based on the analysts' interest.}
The results of i-cNRL heavily depend on the parameters used for each embedding step. 
For example, changing a value of the contrast parameter $\ContParam$ might reveal different unique characteristics in $\GraphT$. 
For analysts with advanced knowledge on NRL and CL, the framework should provide abilities for interactively tuning the i-cNRL results based on their needs.

\begin{figure}[tb]
	\centering
	\captionsetup{farskip=0pt}
    \includegraphics[width=1.0\linewidth]{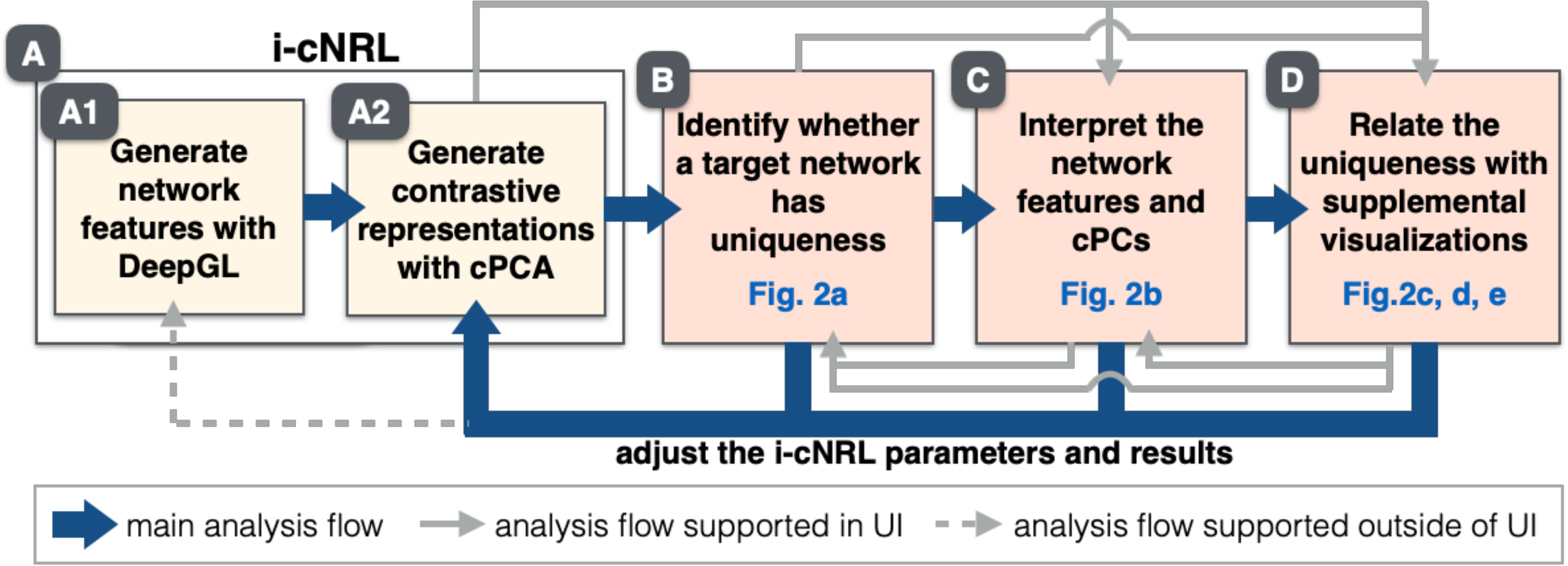}
    \caption{Contrastive network analysis workflow with \name{}.}
	\label{fig:analysis_flow}
\end{figure}

\section{Framework Overview}

Grounded by the \textbf{DIIF} design considerations, we develop \name{} which augments the back-end i-cNRL algorithm with interactive visualization (\autoref{fig:system_overview}), supporting visual contrastive network analysis.

\autoref{fig:analysis_flow} shows a workflow of conducting contrastive network analysis with \name{}.
The workflow starts from (A) generation of the i-cNRL results that includes NRL with DeepGL and CL with cPCA (\autoref{fig:architecture}). 
Afterward, the analyst can first (B) identify whether or not there are any unique characteristics only found in a target network from the contrastive representations visualized by \name{} (\autoref{fig:system_overview}-a).
If such characteristics exist, to understand the uniqueness, the analyst can (C) interpret the network features and cPCs generated by i-cNRL with visualizations in \autoref{fig:system_overview}-b. 
They can also (D) analyze the contrastive representations, network features, and cPCs by relating them with probability distributions (\autoref{fig:system_overview}-c) and laid-out networks (\autoref{fig:system_overview}-d, e). 
Based on findings during the exploration, the analyst might want to adjust the parameters of i-cNRL. 

The above procedure is our expected main analysis workflow as indicated by the thick blue arrows in \autoref{fig:analysis_flow}.
However, the \name{} UI provides the flexibility in the analysis activities, shown by the solid gray arrows in \autoref{fig:analysis_flow}. 
For example, the analyst might want to start to (D) see laid-out networks in order to grasp the topological differences between target and background networks at a glance, and then (B) examine the differences with the contrastive representations. 
Also, such an interactive analysis often requires to go back and forth between different views to validate findings obtained in one view. 

Due to the high computational cost of NRL with DeepGL (e.g., 20 seconds for a network of 6K nodes and 20K links), we decided to support the interactive parameter adjustment only for cPCA. 
After the analyst updates DeepGL's parameters and generates the network features, they can analyze the results with the \name{} UI. 

We have developed \name{} as a web application. 
For the back-end algorithms, we use Python to integrate the existing i-cNRL implementation~\cite{icnrl}.
The front-end UI is implemented with a combination of HTML5, JavaScript, D3~\cite{bostock2011d3}, and WebGL. 
D3 is used for the feature contribution and probability distribution views (\autoref{fig:system_overview}-b, c).
For the other views (\autoref{fig:system_overview}-a, d, e), we utilize WebGL to support efficient rendering and interaction as networks often consist of many nodes and links (e.g., several thousand nodes).
We use WebSocket to communicate between the front- and back-end modules.

\begin{figure}[tb]
	\centering
	\captionsetup{farskip=0pt}
	\captionsetup[subfloat]{width=0.48\linewidth}
	\hspace*{-8pt}
    \subfloat[$\GraphT$: LC-multiple~\cite{reguly2006comprehensive,yu2008high}, \newline
    \hspace*{8pt} $\GraphB$: Combined-AP/MS~\cite{collins2007toward,yu2008high}]{
        \includegraphics[height=0.5\linewidth]{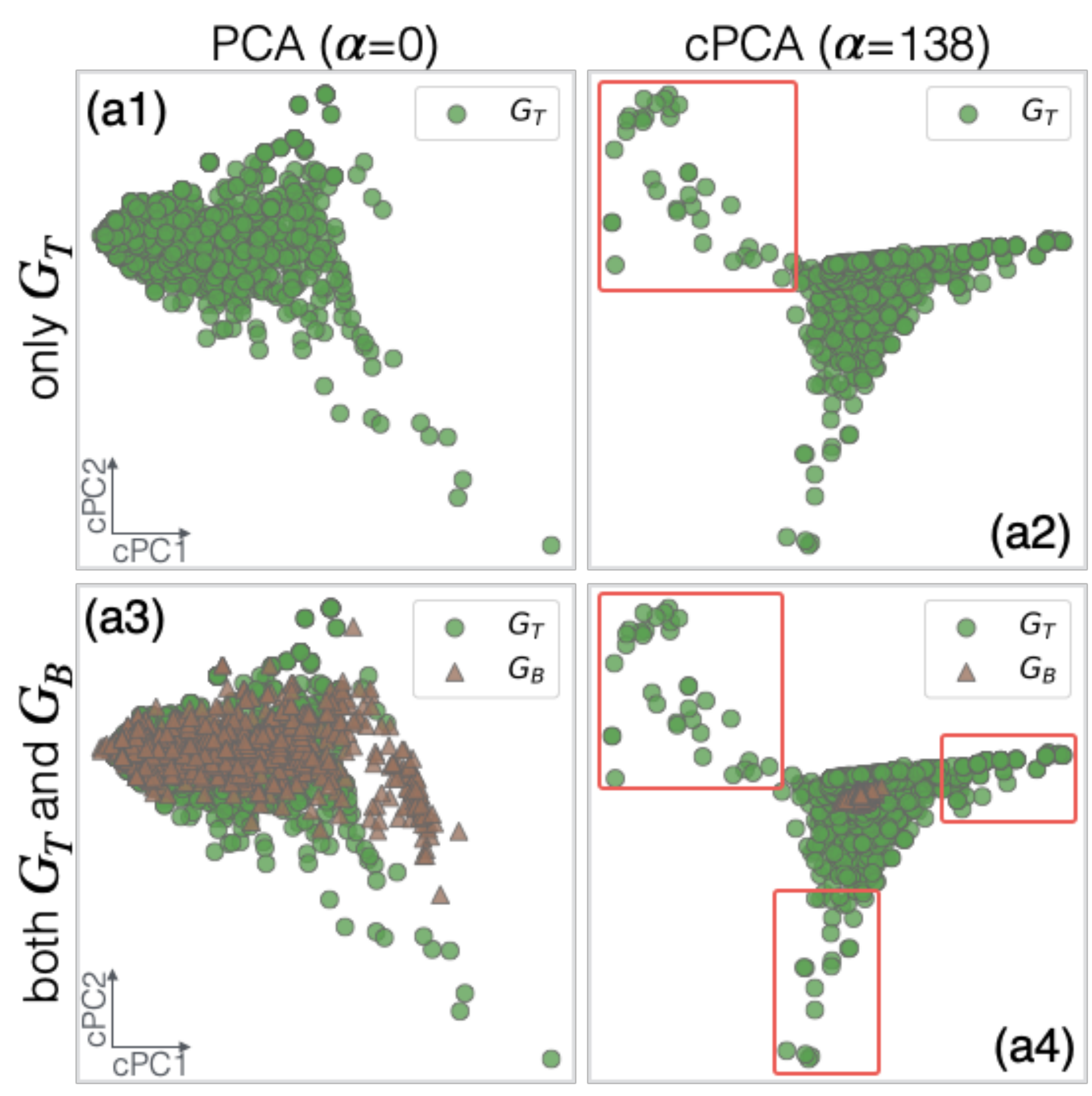}
        \label{fig:vis_cont_repr_a}
    }
    \hspace*{-4pt}
    \subfloat[$\GraphT$: Dolphin~\cite{lusseau2003bottlenose}, $\GraphB$: Karate~\cite{zachary1977information}]{
        \includegraphics[height=0.5\linewidth]{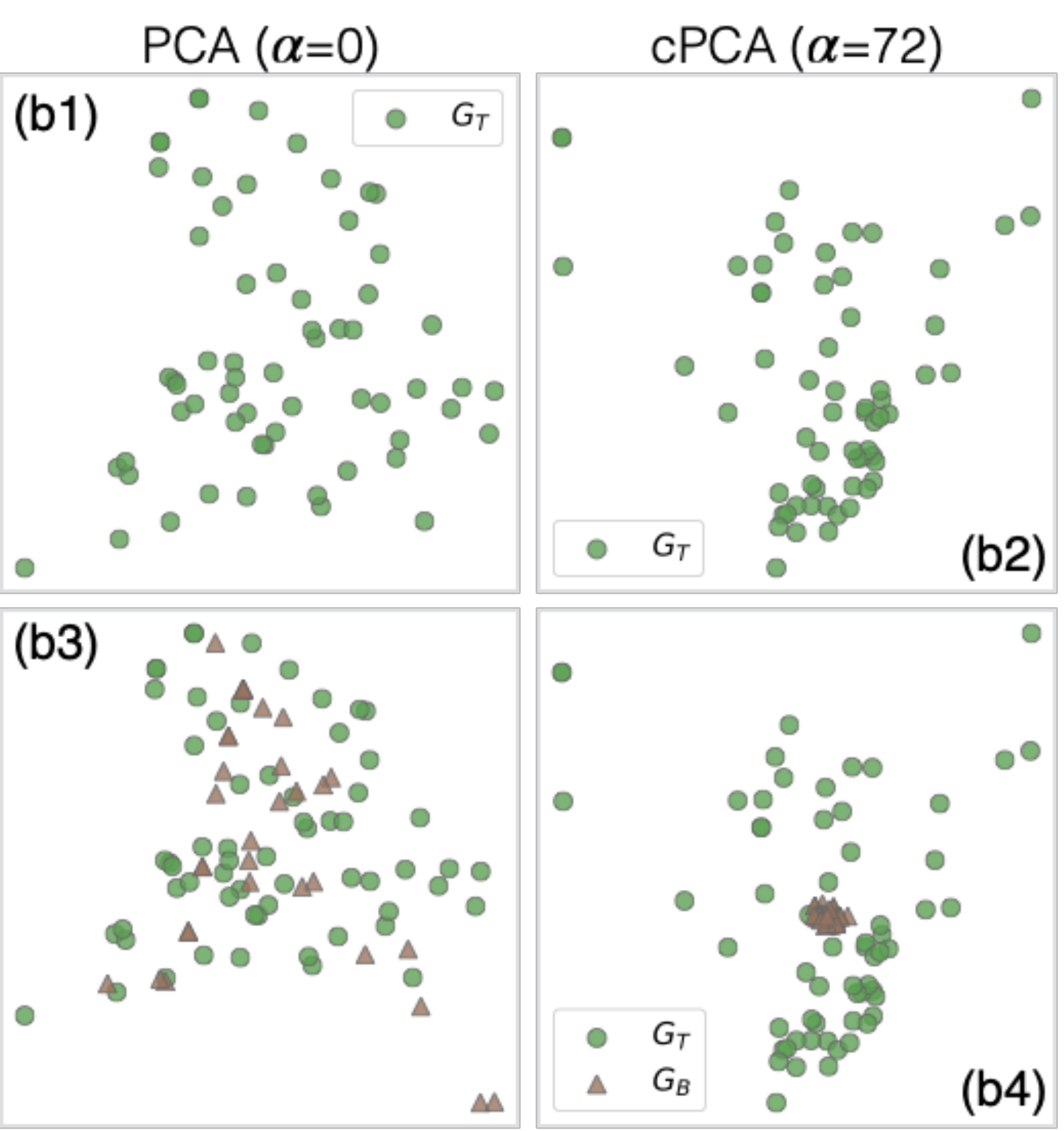}
        \label{fig:vis_cont_repr_b}
    }
    \hspace*{-6pt}
    \caption{Two-dimensional projections based on (contrastive) representations obtained with PCA (when $\ContParam=0$) and cPCA.}
	\label{fig:vis_cont_repr}
\end{figure}

\section{\name{} Visual Interface}
\label{sec:vis}

As shown in \autoref{fig:system_overview}, the \name{} UI consists of four interactively coordinated views, including a contrastive representation view, a feature contribution view, a probability distribution view, and a network layout view, designed with the considerations in \autoref{sec:dc}.  
Here, we describe the views provided by the UI through contrastive network analysis of two social networks, the Dolphin social network~\cite{lusseau2003bottlenose} as $\GraphT$ and Zachary's karate club network~\cite{zachary1977information} as $\GraphB$.
A demonstration video of the interface is available at our online site~\cite{supp}.

\subsection{Visualization of Contrastive Representations}

With the results generated by i-cNRL, the first step of our analysis workflow (\autoref{fig:analysis_flow}-A), 
\name{}'s contrastive representation view (\autoref{fig:system_overview}-a) visualizes the results to reveal whether or not there is uniqueness in the target network compared to the background network, serving as the following step (\autoref{fig:analysis_flow}-B, \textbf{DC1-Discovery}).

\textbf{Visual Identification of Target Network's Uniqueness.}
Similar to existing works~\cite{abid2018exploring,abid2019contrastive,dirie2019contrastive,severson2019unsupervised}, a potential solution is comparing the results of ordinary PCA and cPCA. 
For example, given the two protein interaction networks, LC-multiple~\cite{reguly2006comprehensive,yu2008high} and Combined-AP/MS~\cite{collins2007toward,yu2008high}, \autoref{fig:vis_cont_repr}-a1, a2 show contrastive representations $\ContReprT$ generated with i-cNRL using the contrastive parameter $\ContParam=0$ (PCA) and $\ContParam=138$ (cPCA), respectively.
In \autoref{fig:vis_cont_repr}-a2, comparing with \autoref{fig:vis_cont_repr}-a1, we can see the emergence of a new cluster, as annotated with the red rectangle.  
It indicates that cPCA successfully finds directions (i.e., cPCs) where $\GraphT$ has a higher variance than $\GraphB$ (i.e., the uniqueness). 
However, in many cases, it is difficult to see clear pattern differences between the results of PCA and cPCA, as shown in \autoref{fig:vis_cont_repr}-b1, b2 with the networks of dolphins~\cite{lusseau2003bottlenose} as $\GraphT$ and Karate club members~\cite{zachary1977information} as $\GraphB$. 

The problem is mainly because we do not know how nodes in a background dataset distribute in the embedding space generated by CL.
Thus, we introduce a method that plots the contrastive representations of target and background datasets, $\ContReprT$ and $\ContReprB$, together.
As shown in \autoref{fig:vis_cont_repr}-a3, a4, b3, b4, $\ContReprT$ and $\ContReprB$ are visualized as green circles and brown triangles, respectively.

When a network has high variance in the embedded space, its nodes are widely distributed along cPCs. 
Thus, the uniqueness of a target network $\GraphT$ can be identified by comparing the scatteredness of nodes in $\ContReprT$ and $\ContReprB$.
As shown in \autoref{fig:vis_cont_repr}-a4, b4, cPCA reveals that $\ContReprT$ has much higher scatteredness than $\ContReprB$.
Moreover, we can easily grasp which parts of a target network have strong uniqueness.
Similar to other representation learning methods (e.g., PCA and MDS~\cite{torgerson1952}), a distance in the embedding space of cPCA represents a dissimilarity between nodes.  
Thus, when the target network nodes are highly unique, they are placed far away from the nodes in the background network (e.g., the nodes in the red boxes of \autoref{fig:vis_cont_repr}-a4).

\begin{figure}[tb]
	\centering
	\captionsetup{farskip=0pt}
	\captionsetup[subfloat]{width=0.33\linewidth}
	\hspace*{-5pt}
	\subfloat[Highlighting of $\GraphT$]{
        \includegraphics[height=0.33\linewidth]{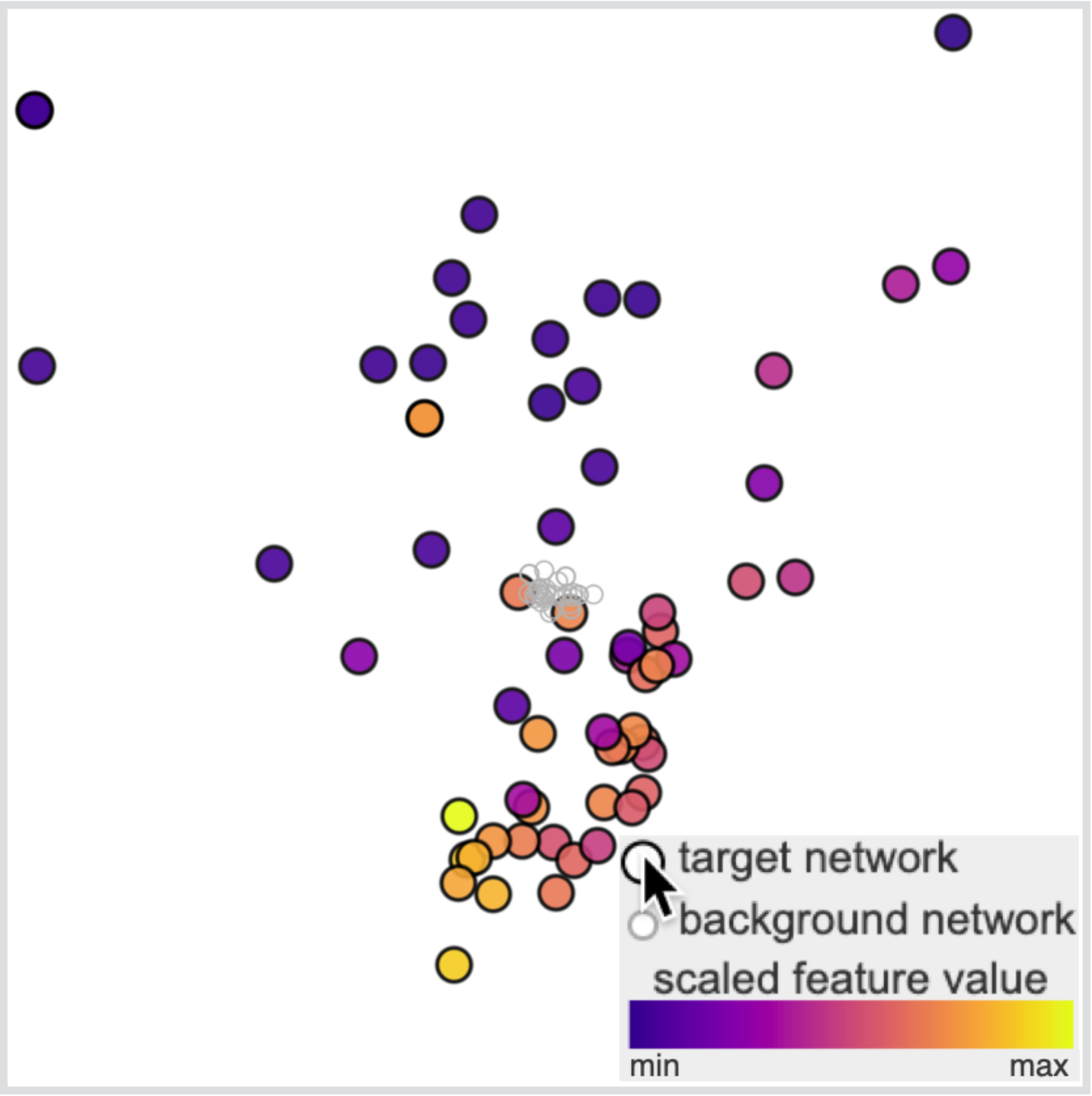}
        \label{fig:selection_tg}
    }
    \hspace*{-5pt}
    \subfloat[Highlighting of $\GraphB$]{
        \includegraphics[height=0.33\linewidth]{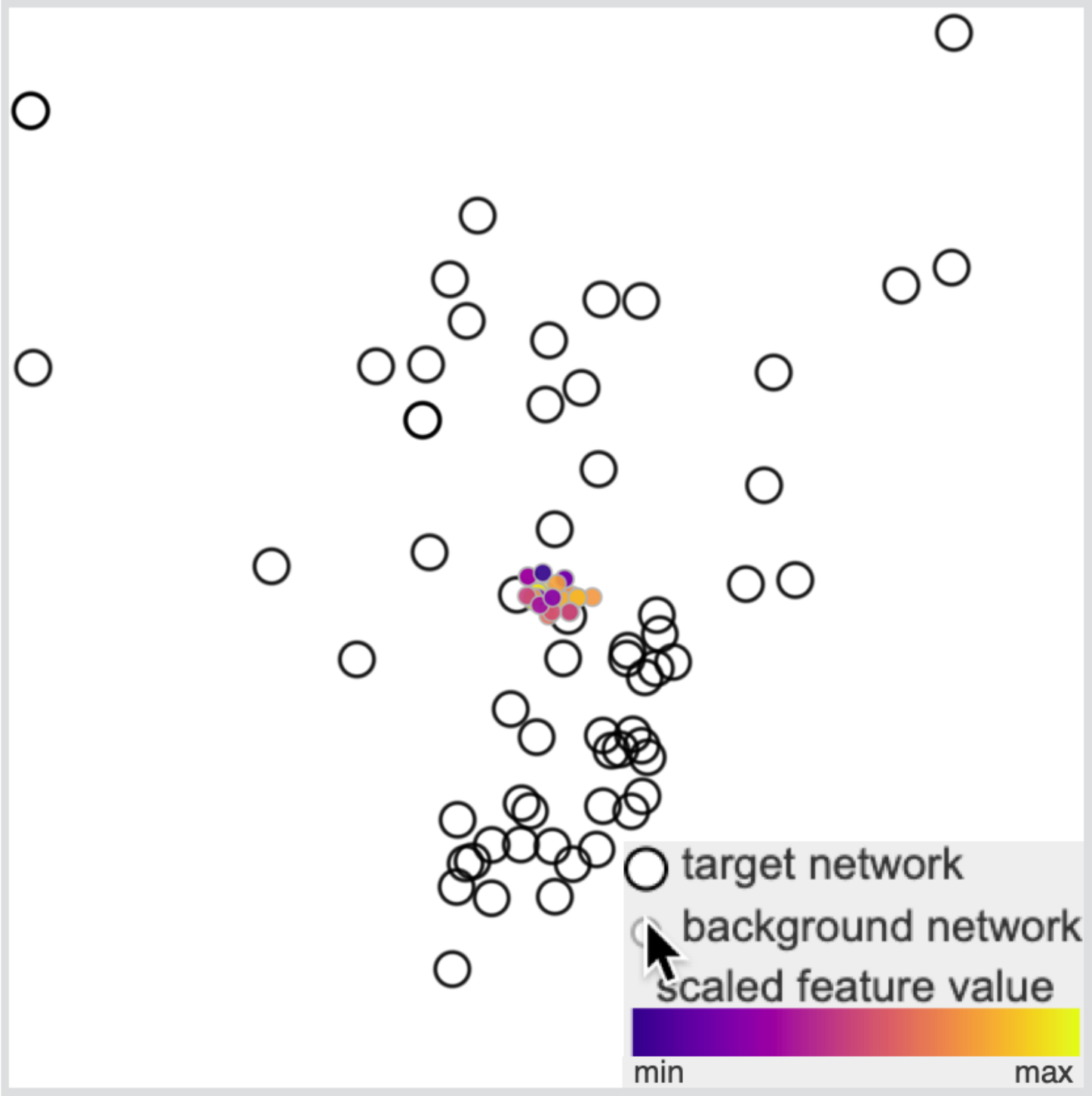}
        \label{fig:selection_bg}
    }
    \hspace*{-5pt}
    \subfloat[Selection with lasso]{
        \includegraphics[height=0.33\linewidth]{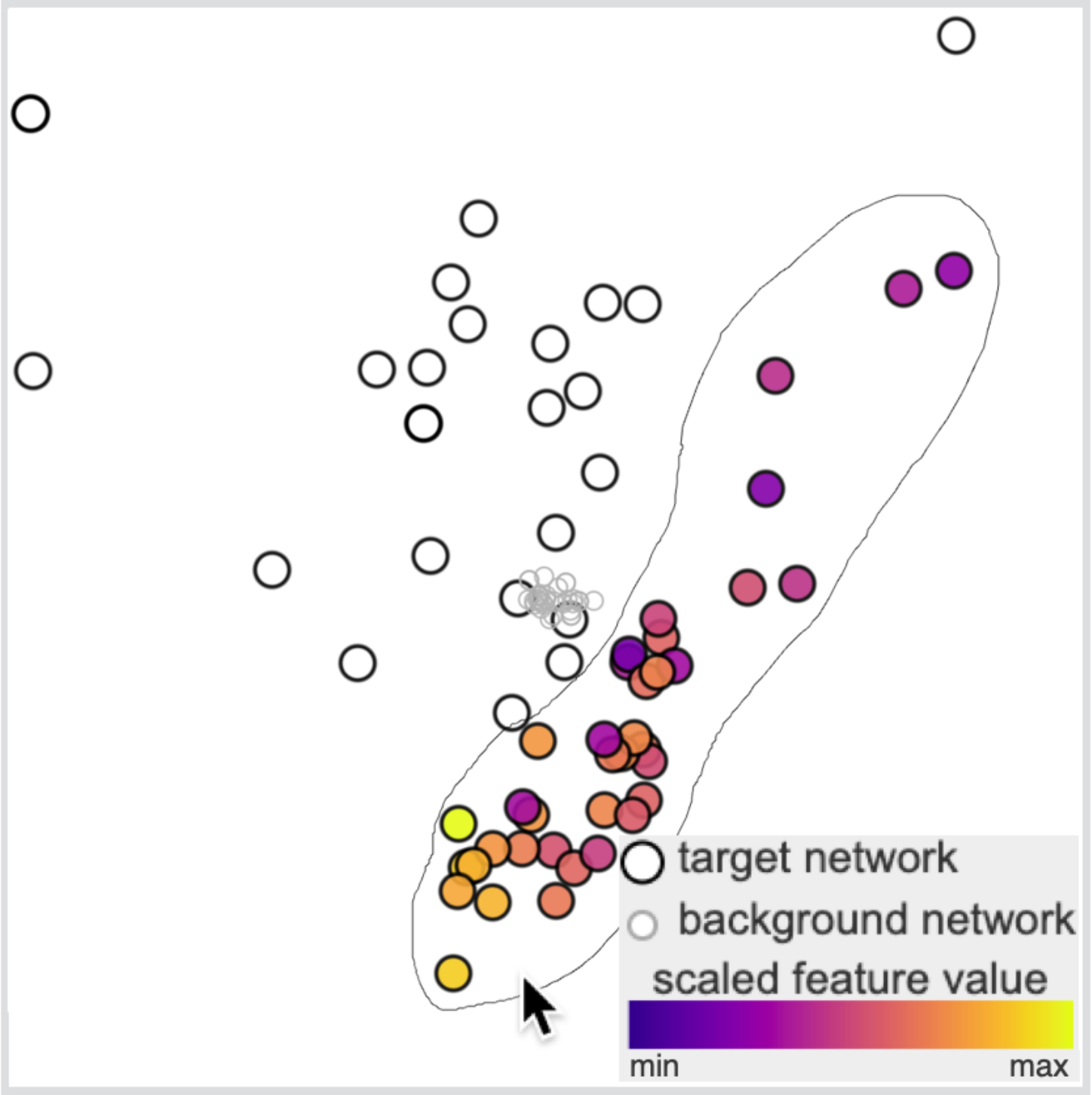}
        \label{fig:selection_lasso}
    }
    \hspace*{-5pt}
    \caption{Node highlighting and selection supported in the contrastive representation view.}
	\label{fig:selection}
\end{figure}

\textbf{Integration into \name{}.}
We employ the above visualization as the contrastive representation view of \name{} (\autoref{fig:system_overview}-a), where the values of a network feature selected in the feature contribution view (see \autoref{fig:system_overview}-b and \autoref{sec:fc_view}) are colorcoded with a purple-yellow scheme~\cite{mplcolormaps}. 
To encode nodes in target and background networks, we first explored different shapes, including circle, triangle, and squares; however, circles and squares are hard to distinguish and triangles require much higher rendering cost with WebGL than circles and squares. 
We then used circles with different sizes and borders, with larger and black-border circles for the target network and smaller and gray-border circles for the background network.
Moreover, the analyst can highlight the target or the background network by hovering over the corresponding legend as shown in \autoref{fig:selection}-a, b. 
The contrastive representation view also provides fundamental interactions, such as zooming, panning, and lasso selection (\autoref{fig:selection}-c). 
From the different scatteredness of $\GraphT$ and $\GraphB$ nodes in \autoref{fig:system_overview}-a, we can decide that there exists uniqueness in the Dolphin network.

\subsection{Interpretation of Network Features and cPCs}
\label{sec:fc_view}

With the above observation from the contrastive representation view, we move on to interpret the network features and cPCs (\autoref{fig:analysis_flow}-C, \textbf{DC2-Interpretability}) with the feature contribution view (\autoref{fig:system_overview}-b).

\begin{figure}[tb]
	\centering
	\captionsetup{farskip=0pt}
	\subfloat[Notations in DeepGL.]{
        \includegraphics[height=0.126\linewidth]{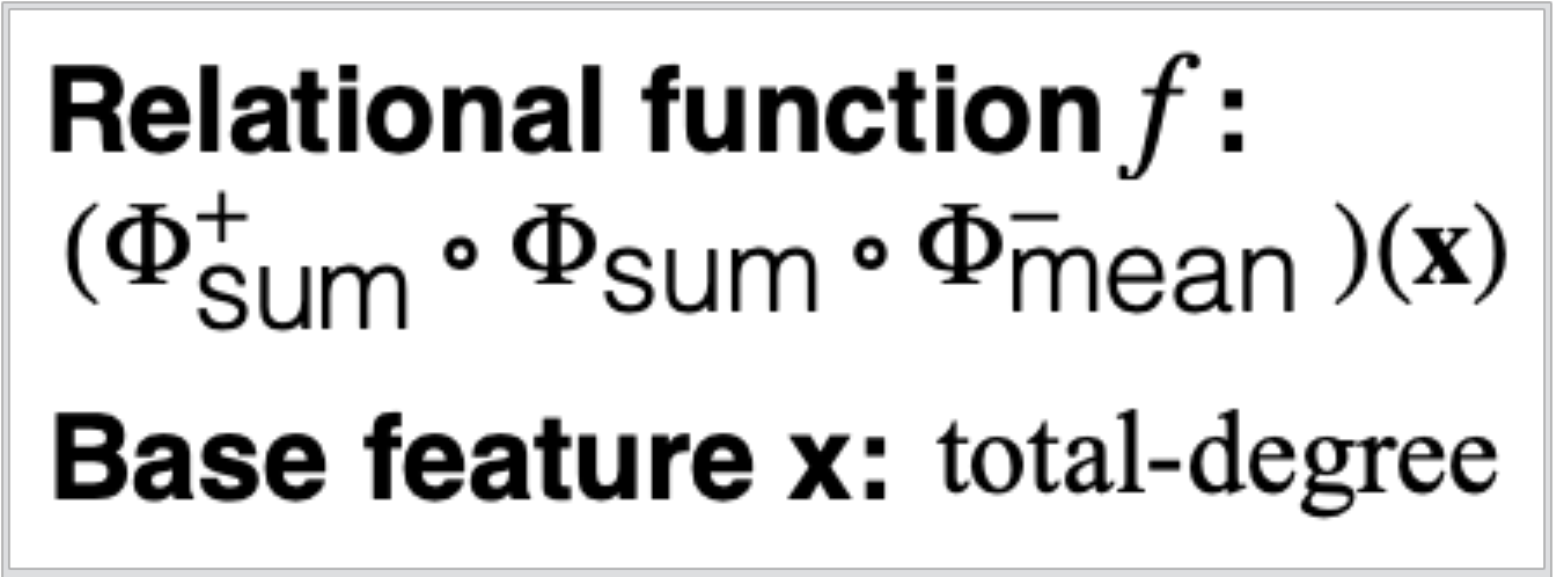}
        \label{fig:feat_repr_deepgl}
    }
    \subfloat[Visual representations in \name{}.]{
        \includegraphics[height=0.126\linewidth]{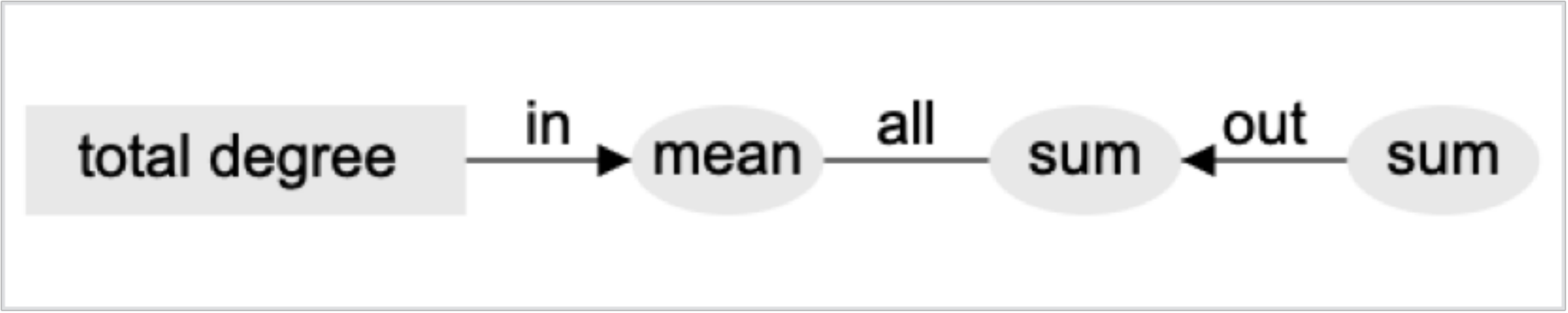}
        \label{fig:feat_repr_contrana}
    }\vspace{4pt}\\
    \subfloat[Computation of the feature $(\RelFeatOpe{+}{\Sum} \circ \RelFeatOpe{}{\Sum} \circ \RelFeatOpe{-}{\Mean})(\BaseFeat)$, where $\BaseFeat$ is \f{total-degree}.]{
        \includegraphics[width=0.99\linewidth]{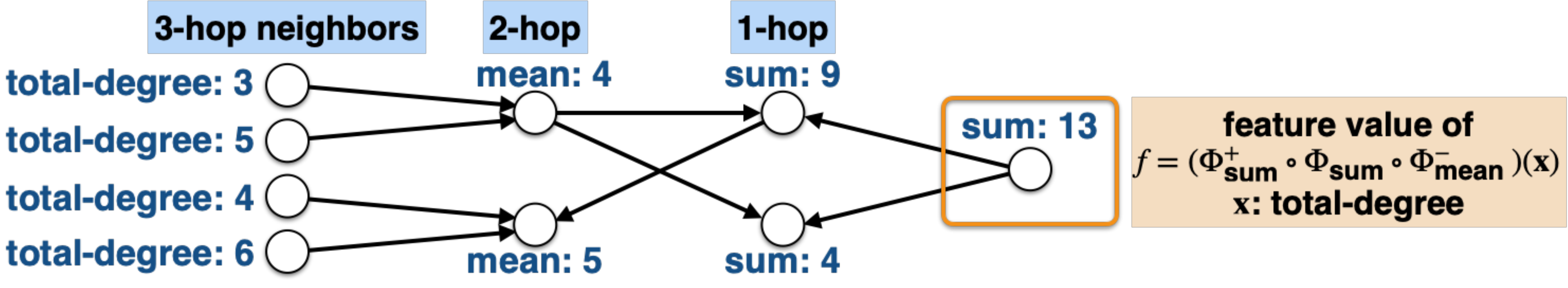}
        \label{fig:feat_repr_computation}
    }
    \caption{Representations of network features in DeepGL and \name{}. Here, as an example, we use a complex feature (consisting of three relational feature operators) that DeepGL may produce. (a) and (b) represent the same feature: \textit{the sum of out-neighbors of the sum of all-neighbors of the mean of in-neighbors of \f{total-degrees}}. (c) shows an example of the computational flow of this feature. In (c), the circles and arrows represent nodes and directed links of a network.}
	\label{fig:feat_repr}
\end{figure}

\textbf{Visual Representation of Network Features.}
The left part of the feature contribution view lists all the network features generated by DeepGL. 
They usually consist of a few relational feature operators (RFOs), which are represented with mathematical notations (\autoref{fig:feat_repr}-a).
However, it is difficult for analysts to interpret features with such notations. 
We thus design an intuitive visual representation of the features (\autoref{fig:feat_repr}-b).

A network feature learned by DeepGL consists of the base feature (e.g., \f{total-degree}), summary measures (e.g., mean), and neighbor types (e.g., in-neighbors). 
We use a gray rectangle and an ellipse with text labels to denote a base feature and a summary measure, respectively.
Then, we connect them with a line and, to indicate the neighbor type, annotate with a text label (\textit{in}, \textit{out}, or \textit{all}).
Also, for in- and out-neighbors, we use an arrowhead to indicate the direction.
Lastly, we order them from left to right based on the computational flow to obtain the feature value. 
The resultant representation in \autoref{fig:feat_repr}-b visually summarizes the neighborhood relationships and the computational flow, which is further explained in \autoref{fig:feat_repr}-c. 

\textbf{Visualization of cPC Loadings.}
The right part of the feature contribution view visualizes cPC loadings~\cite{fujiwara2020supporting} described in \autoref{sec:icnrl} as a heatmap.
Each row and column correspond to a network feature and cPC, respectively.  
Similar to Fujiwara \etal's work~\cite{fujiwara2020supporting}, we generate scaled cPC loadings (or \textit{feature contributions}) between $[-1, 1]$ by dividing each cPC's loadings by their maximum absolute value.
Then, we encode the scaled cPC loadings with a brown-to-blue-green diverging colormap~\cite{palettable,harrower2003colorbrewer}. 
The magnitude of the loading represents how strongly a feature contributes to the corresponding cPC.
For example, the feature at the eighth row in \autoref{fig:system_overview}-b (\f{F8}: the mean of all-neighbors' \f{eigenvector centralities}~\cite{newman2018networks}), has the most influence on cPC1.
Also, the sign of the loading indicates the contributed direction along the cPC ($+$: positive; $-$: negative).
For example, in \autoref{fig:system_overview}-a where each node is colored by \f{F8}, we can see that the feature values of $\GraphT$ generally vary from low to high along the positive $x$-direction. 

By default, \name{} automatically selects the feature that most strongly contributes to cPC1 (e.g., \f{F8} in \autoref{fig:system_overview}-b) and highlights the corresponding row in yellow. 
The analyst can select a different feature, and all other views are updated based on the selected feature (e.g., node colors in the contrastive representation view).

By using the contrastive representation and feature contribution views together, we discover that the uniqueness of the Dolphin network $\GraphT$ highly relates to \f{F8}. 
From the nodes colored by the feature values (\autoref{fig:system_overview}-a), we can see that the nodes around the top-left have low values while the nodes around the bottom-right tend to have higher values.

\subsection{Relating to Common Network Visualizations}

With above results, we further analyze the uniqueness by relating \f{F8} to common network visualizations (\autoref{fig:analysis_flow}-D). 
\name{} provides two perspectives for network analysis (\textbf{DC3-Intuitiveness}): probability distributions and laid-out networks~\cite{barabasi2016network}. 
Probability distributions are often used to compare the distributions of target and background networks' centralities (e.g., whether the degree distribution follows the power law~\cite{barabasi2016network}), and laid-out networks are helpful for viewing the topological differences (e.g., whether multiple communities exist).

\textbf{Linking with Probability Distributions.}
The probability distribution view (\autoref{fig:system_overview}-c) shows the distributions of the selected feature values in the feature contribution view (i.e., \f{F8} in \autoref{fig:system_overview}-b), for target and background networks. 
Its $x$- and $y$-coordinates represent a (scaled) feature value and its probability (or relative frequency), respectively. 
Both logarithmic and linear scales for the $y$-coordinate are supported.
We colorcode the probability distribution lines with the same colors used for the node borders in the contrastive representation view (i.e., black: target network, gray: background network).

\textbf{Linking with Network Layouts.}
The network layout view in \autoref{fig:system_overview}-d, e visualizes laid-out target and background networks, with the scalable force-directed placement~\cite{hu2005efficient}. 
Same as the contrastive representation view, each node is colored based on the selected feature in the feature contribution view (e.g., \f{F8} in \autoref{fig:system_overview}-b) and outlined in black (target network) or gray (background network).
The network layout view also supports several basic interactions such as zooming, panning, and lasso selection, and is fully linked with other views.
For example, by reviewing \autoref{fig:system_overview}-a, b, d together, we notice that the two node groups found previously (i.e., nodes with small and high \f{F8} values, placed around the top-left and bottom-right in \autoref{fig:system_overview}-a) seem to correspond to distinct communities at the bottom-left and top-right in \autoref{fig:system_overview}-d. 
This can be confirmed by performing a lasso selection on the nodes in \autoref{fig:system_overview}-a, as demonstrated in \autoref{fig:selection}-c. 

\begin{figure}[tb]
	\centering
	\captionsetup{farskip=0pt}
    \includegraphics[width=1.0\linewidth]{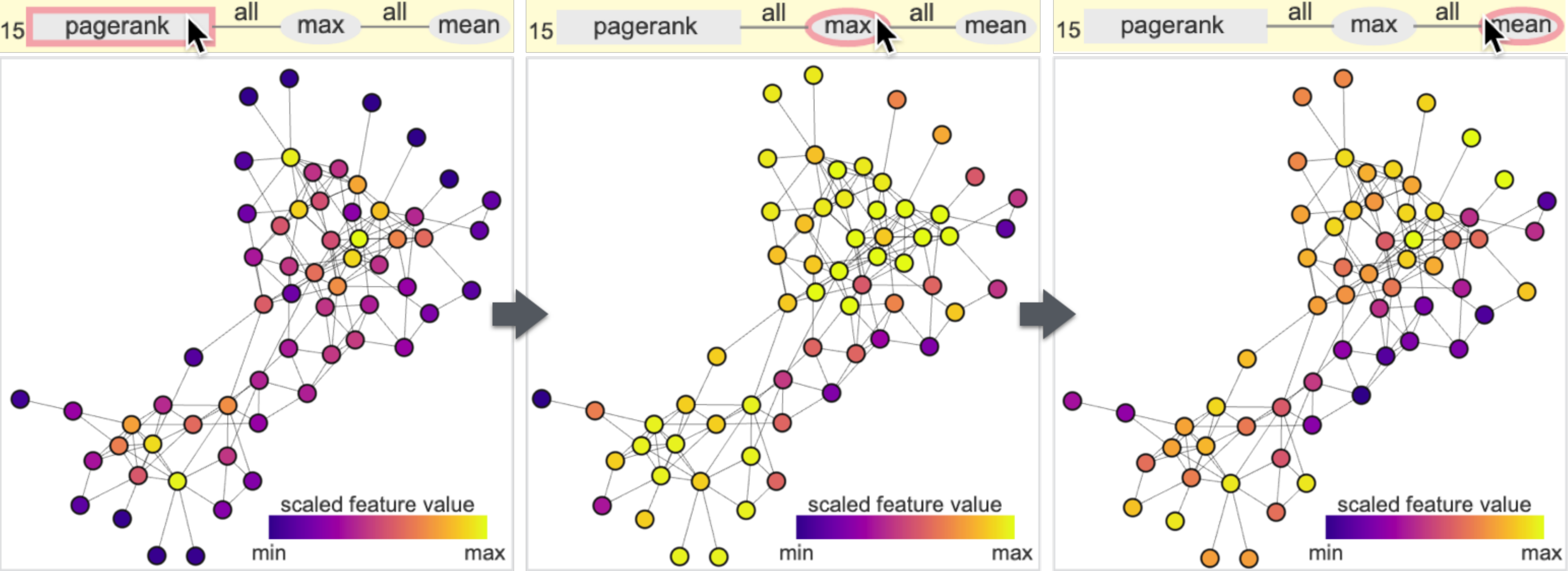}
    \caption{Visualization of intermediate computational results of feature \f{F15}.}
	\label{fig:step_by_step}
\end{figure}

\textbf{Understanding Complicated Network Features.}
The linkings above can be utilized to further help understand the network feature that consists of multiple RFOs. 
As shown in \autoref{fig:step_by_step}, by hovering over either the base feature or summary measure in the feature contribution view, the network layout view and the contrastive representation view show the intermediate computational results of the feature values. 
For instance, \autoref{fig:step_by_step} (from left to right) visualizes \f{PageRank} values of $\GraphT$'s nodes, the maximum of all-neighbors of \f{PageRank} values, and the mean of all-neighbors of them.  
Thus, the analyst can visually understand how the base feature values spread across the neighbors and how the final network feature values are derived. 

Through the analysis from \autoref{fig:analysis_flow}-A to D, we can conclude that the Dolphin network $\GraphT$ has unique characteristics relative to the Karate network $\GraphB$. 
The uniqueness highly relates to \f{F8: eigenvector centralities} of each node's neighbors, and it clearly reveals the separation of the two communities in $\GraphT$, which cannot be seen in $\GraphB$.

\subsection{Refinement of Contrastive Representations}
\label{sec:refine_cr}

\begin{figure}[tb]
	\centering
	\captionsetup{farskip=0pt}
    \subfloat[Without sign adjustment.]{
        \includegraphics[width=0.99\linewidth]{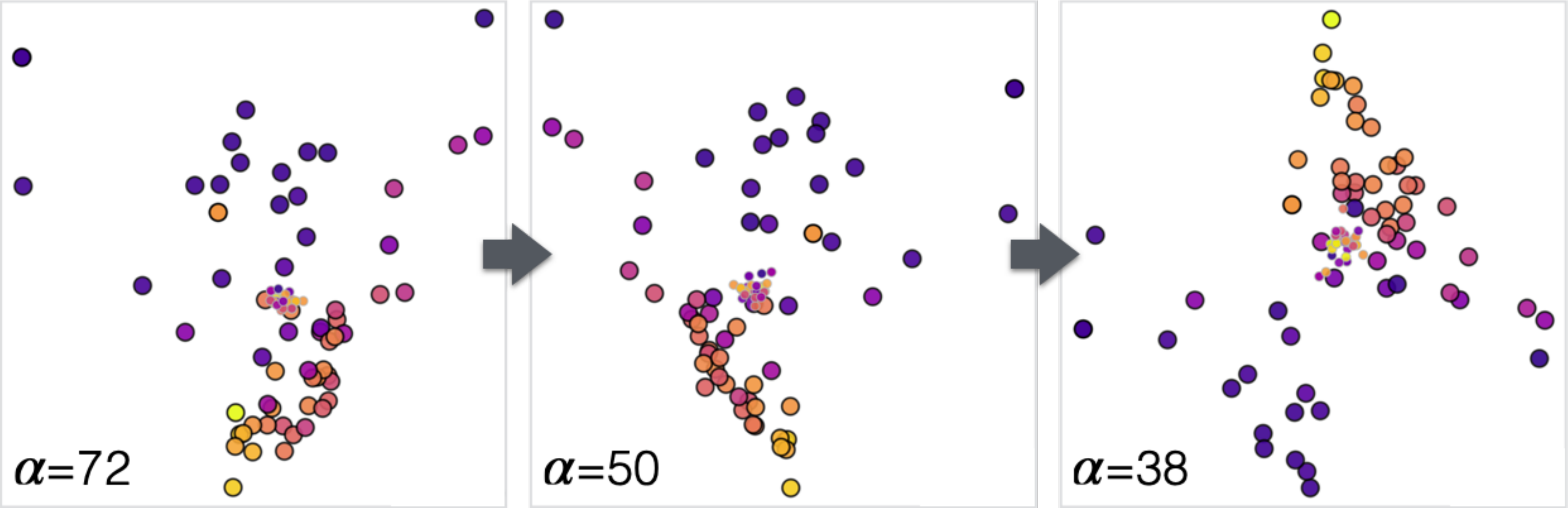}
        \label{fig:alpha_adj_without}
    }\vspace{2pt}\\
    \subfloat[With sign adjustment.]{
        \includegraphics[width=0.99\linewidth]{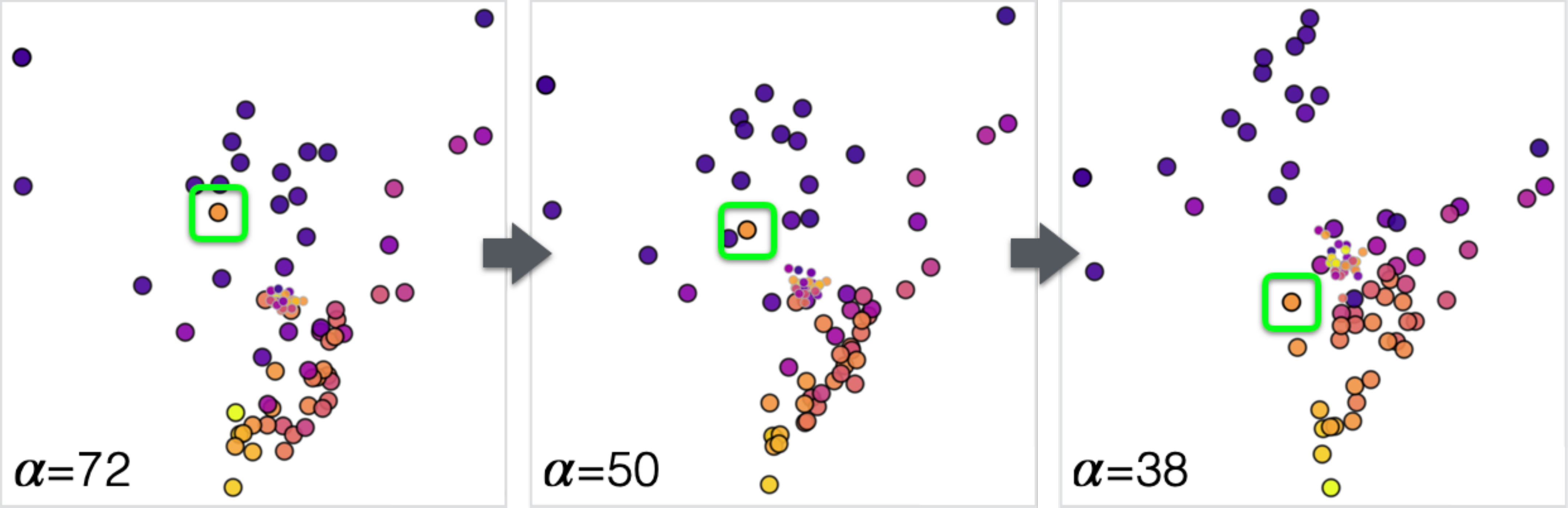}
        \label{fig:alpha_adj_with}
    }
    \caption{Changes of node projection by updating contrastive parameter $\ContParam$ (a) without and (b) with sign adjustment. Corresponding animations are available online~\cite{supp}.
    }
	\label{fig:alpha_adj}
\end{figure}

The cPCA used in i-cNRL automatically selects the contrastive parameter $\ContParam$ and computes cPCs to generate the optimized contrastive representations, i.e., maximizing the variation in $\FeatMatT$ while simultaneously minimizing the variation in $\FeatMatB$~\cite{icnrl} (\autoref{fig:architecture}).
However, the analyst may want to loosen or strengthen the reduction of the variation of $\FeatMatB$ in order to elucidate the found patterns or discover different patterns.
For example, around the top-left in \autoref{fig:system_overview}-a, an orange node, with a high value of \f{F8}, is mixed up with the nodes with lower values (as annotated in the green box in \autoref{fig:alpha_adj}-b). 
Also, the resultant cPCs might not apt to interpret visually found patterns.
For example, in \autoref{fig:system_overview}-a, the value of \f{F8} tends to increase along the diagonal line, but not along cPC1 (the $x$-axis).
To handle such cases, \name{} supports interactive adjustments of $\ContParam$ and cPCs (\textbf{DC4-Flexibility}).

\textbf{Adjustment of Contrastive Parameter.}
\name{} allows the analyst to interactively change the contrastive parameter $\alpha$ with a range slider (\autoref{fig:system_overview}-f), based on the efficiency of cPCA (e.g., the completion time is less than 3\si{\milli\second} for 10,000 nodes with 10 features~\cite{fujiwara2020supporting}).
However, the update of $\alpha$ in cPCA causes an arbitrary sign flipping for each cPC, similar to PCA~\cite{turkay2017designing,fujiwara2019incremental}. 
\autoref{fig:alpha_adj}-a shows an example of the flipping along both horizontal and vertical directions when $\alpha$ is changed, making it difficult to follow.

To address this, we employ a similar solution used for PCA~\cite{turkay2017designing}.
For each of cPC1 and cPC2, we compute the cosine similarity between the coordinates of all nodes before and after the update; then if the similarity is negative, we flip the sign generated by cPCA. 
\autoref{fig:alpha_adj}-b shows the result with the sign adjustment. 
As $\ContParam$ decreases to $38$, the orange node annotated with the green rectangle moves toward the right-bottom and the separation of nodes with low (purple) and higher values (pink, orange, and yellow) becomes more salient. 

\begin{figure}[tb]
	\centering
	\captionsetup{farskip=0pt}
    \includegraphics[width=0.99\linewidth]{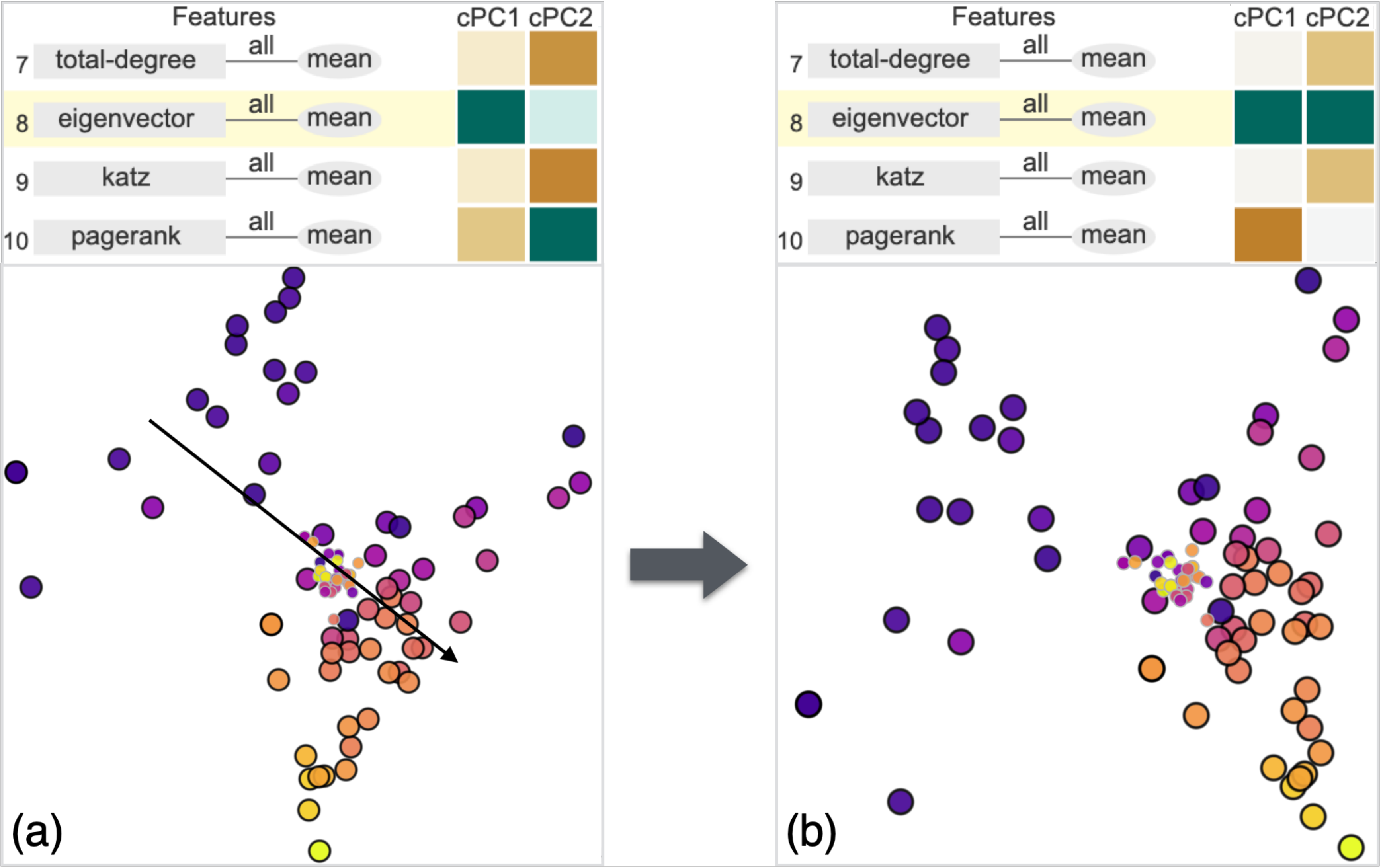}
    \caption{Adjustment of the cPCs. (a) and (b) show feature contributions of Features 7--10 and the contrastive representation view before and after the adjustment, respectively.}
	\label{fig:cpc_adj}
\end{figure}

\textbf{Adjustment of Contrastive Principal Components.}
We introduce an interactive method for customizing cPCs, which can be used for both PCA and cPCA. 
First, in the contrastive representation view, a preferable axis direction for cPC1 can be drawn as a straight line (\autoref{fig:cpc_adj}-a).
We then rotate the projection based on the angle between the drawn line and cPC1 (\autoref{fig:cpc_adj}-b). 
As a result, we also need to update the cPC loadings shown in \autoref{fig:system_overview}-b.
Similar to the rotation in ordinary PCA~\cite{richman1986rotation}, the cPC loadings can be obtained by simply multiplying a rotation matrix with the above user-defined angle.
For example, \autoref{fig:cpc_adj}-a, b show a subset of the cPC loadings corresponding to \f{F7-10} before and after the rotation. 
We can see that \f{F8} has a strong contribution to both cPC1 and cPC2 and \f{F10} has a stronger contribution to cPC1 than before.

Note that we can also use a method developed by Kwon \etal~\cite{kwon2016axisketcher} for general scatterplots, including cPCA projection results.
It generates new axes based on the user-drawn freeform line over the plot and nonlinear transformation.
However, we use the above linear transformation, so that we can update cPC loadings, which are important to interpret the result of cPCA. 

\section{Case Studies}
\label{sec:case_study}

In \autoref{sec:vis}, we have shown the effectiveness of \name{} through an example of comparing two social networks.
Here, we demonstrate two additional case studies, including an evaluation of a network model and a comparison of protein interaction networks.

\begin{figure}[tb]
	\centering
	\captionsetup{farskip=0pt}
    \includegraphics[width=1.0\linewidth]{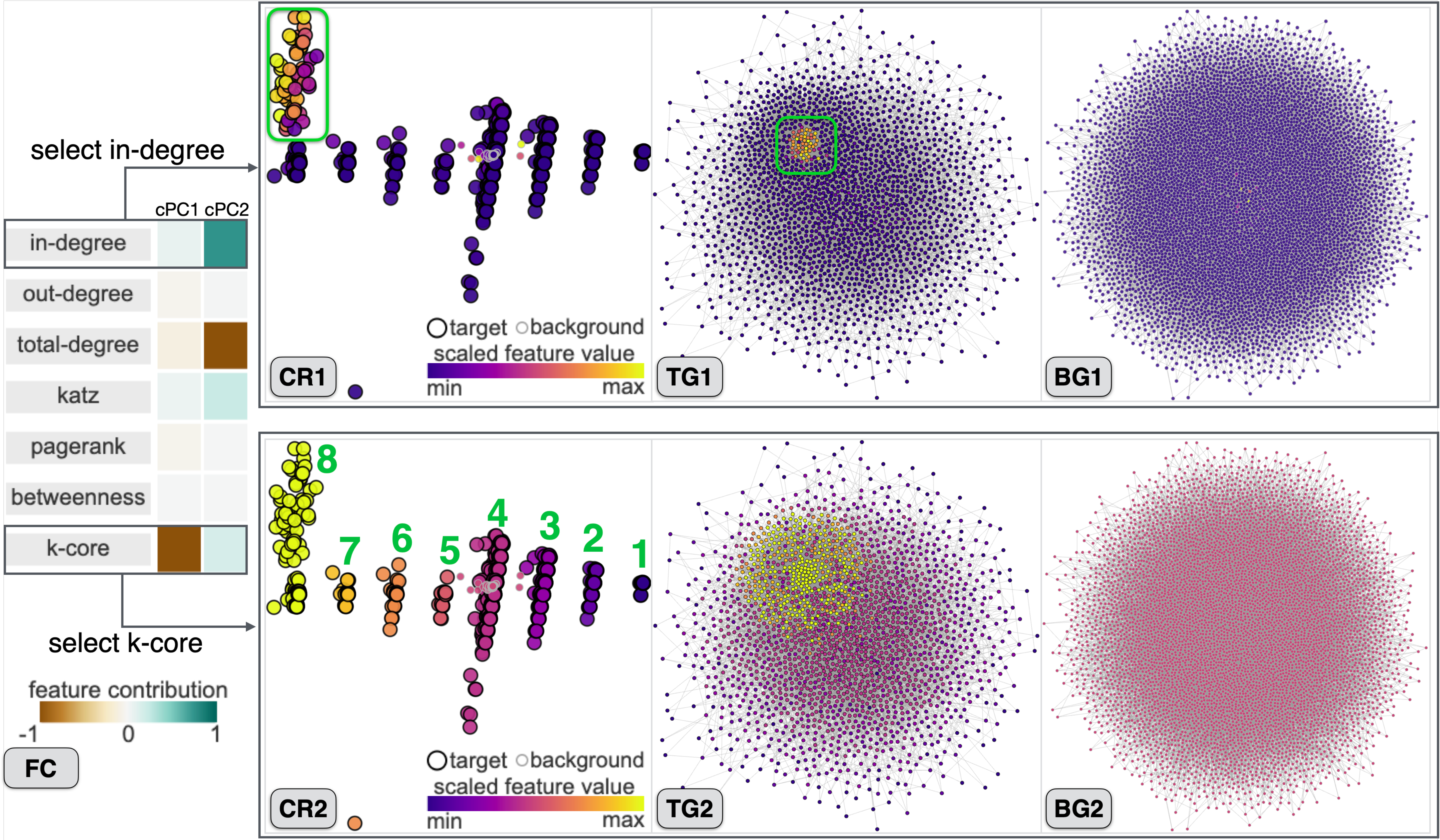}
    \caption{Case study 1. (FC) shows a subset of network features visualized in the feature contribution view (note: a full set of them includes a feature consisting of multiple RFOs).
    (CR1, TG1, BG1) show the contrastive representation, target network layout, background network layout views after selecting \f{in-degree}, respectively. 
    Similarly, (CR2, TG2, BG2) are the results after selecting \f{$k$-core}. 
    As a default, \name{} selects \f{$k$-core} as the most contributed feature to cPC1. }
	\label{fig:cs1}
\end{figure}

\subsection{Study 1: Evaluation of a Network Model}
\label{sec:cs1}

Network modeling is essential to simulate and understand real-world networks (e.g., how they grow and shrink).
It also can be used to perform what-if analysis (e.g., what elimination of a hub node will cause), as well as to generate synthetic datasets~\cite{goldenberg2010survey}. 
\name{} can help evaluate network models by comparing them with real-world networks. 
In this case study, we focus on peer-to-peer (P2P) networks, where a precise network model is essential for analyzing the robustness of a P2P network~\cite{liu2009efficient}.
P2P networks are often scale-free~\cite{liu2009efficient}; thus, we use the Price's model~\cite{newman2018networks} as an evaluation target. 
As listed in \autoref{table:network_info} (see \autoref{sec:user_study}), to identify which characteristics the Price's model does not simulate well, we set a real-world P2P network (p2p-Gnutella08) as $\GraphT$, and the network generated with the Price's model (Price 2) as $\GraphB$.

As shown in \autoref{fig:cs1}-CR2, the contrastive representation view indicates the differences in the node distributions of $\GraphT$ and $\GraphB$. 
The P2P network ($\GraphT$) has clear groups of nodes, unlike Price 2 ($\GraphB$).
From \autoref{fig:cs1}-FC, \f{in-degree}, \f{total-degree}, and \f{$k$-core} are identified as main contribution features.
After selecting \f{in-degree} in \autoref{fig:cs1}-FC to review the related information with the other views (\autoref{fig:cs1}-CR1, TG1, BG1), we notice that $\GraphT$ has a region where nodes have a much higher \f{in-degree} than the other nodes, as annotated with the green boxes in \autoref{fig:cs1}-CR1, TG1.
Similar findings appear when the \f{total-degree} is selected in \autoref{fig:cs1}-FC. 
As for the \f{$k$-core number} (\autoref{fig:cs1}-CR2, TG2, BG2), there is no obvious difference of this feature in $\GraphB$, but a clear distinction with 8 groups of nodes in $\GraphT$, as annotated in \autoref{fig:cs1}-CR2. 
Moreover, from \autoref{fig:cs1}-FC, we can see that cPC1 is more related to the \f{$k$-core} values and cPC2 is more related to the degree of nodes.
Therefore, unique characteristics in the P2P network have been identified.

Also, the result above reveals more variations of the \f{$k$-core number} in the P2P network $\GraphT$.
The \f{$k$-core number} informs that a node at least connects to other $k$ nodes~\cite{newman2018networks}, indicating that the Price's model presents a significant difference in the network robustness from the P2P network.
This issue arises because the Price's model forms a network by always adding a new node with a fixed number of links; as a result, the generated nodes have a constant \f{$k$-core number}. 
Therefore, to better simulate the P2P network, we should develop or use a model that can generate multiple \f{$k$-core numbers}, such as the dual-Barab\'{a}si-Albert model~\cite{moshiri2018dual} or its extension~\cite{icnrl}.

\begin{figure}[tb]
	\centering
	\captionsetup{farskip=0pt}
    \includegraphics[width=1.0\linewidth]{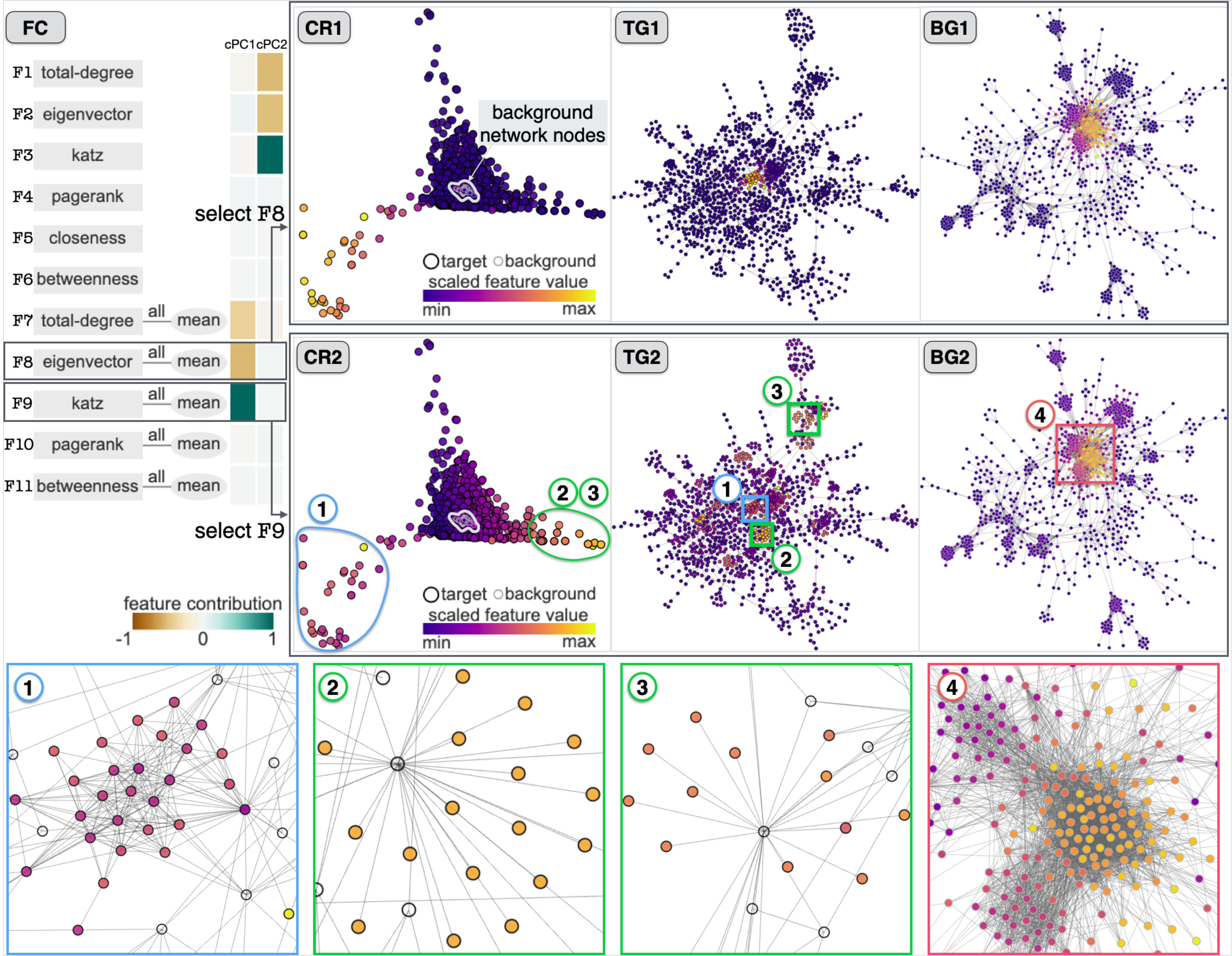}
    \caption{Case study 2. (FC) shows a subset of features in the feature contribution view.
    (CR1, TG1, BG1) show the contrastive representation, target network layout, background network layout views after selecting \f{F8}. 
    Similarly, (CR2, TG2, BG2) are the results after selecting \f{F9}. 
    \textcircled{\small 1}--\textcircled{\small 4} at the bottom show the zoomed regions from TG2 and BG2.}
	\label{fig:cs2}
\end{figure}

\subsection{Study 2: Comparison of Interactome Networks}
\label{sec:cs2}

In this case study, we compare two \textit{interactome} networks, LC-multiple and Combined-AP/MS~\cite{yu2008high}, which represent protein-protein interactions of the yeast \textit{S.~cerevisiae}.
While LC-multiple is the literature-curated (LC) interactome from several \textit{low-throughput} experiments~\cite{yu2008high,reguly2006comprehensive}, Combined-AP/MS is generated through a \textit{high-throughput} approach, specifically, affinity purification/mass spectrometry (AP/MS)~\cite{collins2007toward,yu2008high}. 
Low and high-throughput approaches have different pros and cons in capturing the protein interactions~\cite{yu2008high,wilkins2009hares}, and thus they produce different interactomes.
A comparison of these interactomes is fundamental to assess the quality and characteristics of each approach~\cite{yu2008high}. 
As listed in \autoref{table:network_info}, we set LC-multiple and Combined-AP/MS as $\GraphT$ and $\GraphB$, respectively.

As shown in \autoref{fig:cs2}-CR2, TG2, BG2, \name{} automatically selects \f{F9} (because of its high contribution) and generates related visualizations.
A difference is revealed in the scatteredness of the target and background networks in \autoref{fig:cs2}-CR2. 
We also notice that LC-multiple $\GraphT$ has relatively high feature values towards both left and right directions of cPC1, as annotated with the light blue and green.  
This indicates that cPC1 is not dominantly decided by \f{F9}. 
Therefore, we select the secondary contributed feature \f{F8}, and the results are shown in \autoref{fig:cs2}-CR1, TG1, BG1.  
From \autoref{fig:cs2}-CR1, we can see that only the area annotated light blue in \autoref{fig:cs2}-CR2 has high values of \f{F8}.
\f{F8} and \f{F9} are related to the \f{eigenvector} and \f{Katz centralities}.
Both centralities measure how strongly a node influences other nodes; however, the \f{eigenvector centrality} tends to be high only when a node is in a strongly connected region while the \f{Katz centrality} can be high even when a node is in a weakly connected region~\cite{newman2018networks}.
Thus, we can expect that the nodes annotated with the light blue and green in \autoref{fig:cs2}-CR2 are in strongly and weakly connected regions, respectively.

To visually confirm the above patterns, we select the annotated nodes from \autoref{fig:cs2}-CR2 and zoom into the related regions in \autoref{fig:cs2}-TG2, as detailed by \autoref{fig:cs2}-\textcircled{\small 1}, \textcircled{\small 2}, \textcircled{\small 3}. 
Here we show only two from all the regions related to the nodes annotated with the green color in \autoref{fig:cs2}-CR2, TG2. 
Similarly, in \autoref{fig:cs2}-\textcircled{\small 4}, we show the region where the nodes have high \f{F9} values in \autoref{fig:cs2}-BG2.
We can see that the nodes in \autoref{fig:cs2}-\textcircled{\small 1}, \textcircled{\small 4} are strongly connected, but not in \autoref{fig:cs2}-\textcircled{\small 2}, \textcircled{\small 3}. 
From these observations, we can confirm that the uniqueness is derived from the fact that $\GraphT$ has two different types of nodes linked to high \f{Katz centrality} node(s) in either strongly or weakly connected region, which cannot be seen in $\GraphB$. 
This finding indicates that using LC-multiple or Combined-AP/MS to identify important proteins for \textit{S.~cerevisiae} could reach different conclusions. 
Therefore, additional validation would be needed before deciding their importance based on only one dataset.

\section{Controlled User Study}
\label{sec:user_study}

In addition to the case studies, we conducted a controlled user study to assess the usefulness of \name{} for contrastive network analysis. 
We aimed to answer these research questions: \textbf{(Q1)} Can analysts effectively identify unique characteristics in a target network (compared to a background network), and \textbf{(Q2)} Can analysts properly interpret and explain the found uniqueness? 
We expected that \textbf{Q1} would be primarily addressed by the contrastive representation view, and that all the other coordinated views would help answer \textbf{Q2}.     
We provide the materials used for the study online~\cite{supp}, including the datasets listed in \autoref{table:network_info}, their visualized results with \name{}, and questionnaires.

\subsection{Study Design}
As far as we know, \name{} is the first framework designed for contrastive network analysis, and thus we were not able to find a baseline system to compare against. 
Therefore, we design the following study to evaluate the usability of \name{} in terms of discovering a target network's uniqueness and interpreting it. 

\textbf{Datasets.}
As shown in \autoref{table:network_info}, we generated random networks (Random 1, 2) with Gilbert's random graph~\cite{barabasi2016network} and scale-free networks (Price 1, 2) with the Price's preferential attachment models~\cite{newman2018networks}, as well as used several public datasets. 
We categorized the analysis tasks into three by carefully selecting target and background networks: (a) no uniqueness is in $\GraphT$ (\textit{\# of RFOs} is \texttt{N/A}), (b) the uniqueness in $\GraphT$ can be identified and interpreted with a network feature containing only the base feature (\textit{\# of RFOs} $= 0$), and (c) containing RFOs (\textit{\# of RFOs} $\geq 1$). 
As the number of RFOs increases, a feature becomes more complicated and the task becomes harder. 

\textbf{Participants.}
We recruited 12 participants (4 females and 8 males; aged 18--44) at a local university, with 10 from computer science and 2 from political science. 
There were 1 postdoc-fellow, 10 PhDs, and 1 Master's.
We pre-screened participants to ensure that they have fundamental knowledge of network science.
Their self-reported familiarity with network analysis had the median of 5 (\begin{sparkline}{5} \sparkrectangle 0.0 1.0
    \sparkspike .083 0.000
    \sparkspike .226 0.333
    \sparkspike .369 0.333
    \sparkspike .512 1.000
    \sparkspike .655 1.000
    \sparkspike .797 0.666
    \sparkspike .940 0.666
\end{sparkline}), on a scale of 1 (not familiar) to 7 (use regularly).
Out of 7 network centralities/measures used in the study (i.e., \f{degree, closeness, betweenness, eigenvector, Katz centralities, PageRank,} and \f{$k$-core number}~\cite{newman2018networks}), participants' knowledge of these had the median of 3 (\begin{sparkline}{5} \sparkrectangle 0.0 1.0
    \sparkspike .083 0.25
    \sparkspike .226 0.25
    \sparkspike .369 1.00
    \sparkspike .512 0.50
    \sparkspike .655 0.25
    \sparkspike .797 0.50
    \sparkspike .940 0.25
\end{sparkline}).

\textbf{Apparatus.}
The study was conducted on an iMac (4 GHz Intel Core i7, 16GB 1,600 MHz DDR3) with a 27-inch display ($5,120 \times 2,880$ pixels), connected with an Apple Magic Mouse 2.
The UI was presented with Google Chrome in full-screen mode. 
Because the refinement of contrastive representations (\autoref{sec:refine_cr}) was not relevant to our study tasks, we disabled the related functionalities.

\textbf{Tasks and Design.}
Based on \textbf{Q1} and \textbf{Q2}, given target and background networks, participants were asked to perform comparative analysis using \name{} and complete two subtasks,  \textbf{(ST1)} and \textbf{(ST2)}: \textbf{(ST1)} identifies whether or not the target network has any uniqueness compared to the background network, and \textbf{(ST2)} explains the found uniqueness (if any) or the reason of concluding there is no uniqueness. 
ST1 required a selection from options of \textit{Yes}, \textit{No}, and \textit{I'm not sure}; for ST2, participants were asked to write down their explanation.
We employed a within-subjects design for our study. Each participant completed three comparative network analysis tasks in our main study, using three different pairs of networks (Tasks A, B, and C in \autoref{table:network_info}). The order of tasks was counterbalanced across participants.    

\textbf{Procedure.}
At the beginning, participants provided their demographics and backgrounds on a survey. 
A brief tutorial was then presented including explanations of the definition of the uniqueness, the above 7 network centralities/measures, the usage of \name{}, and 3 concrete analysis examples.
Afterward, participants completed a training session, allowing them to get familiar with \name{} and the task, followed by the real study consisting of three tasks.
The datasets used in the tutorials, training, and study tasks are shown in \autoref{table:network_info}.
Think aloud protocol was used during the training and task sessions. 
They were allowed to ask questions about the \name{} UI and the network centralities and measures. 
No time limit was set for the tasks.
Lastly, participants provided their feedback with the NASA TLX~\cite{Hart1988development}, a questionnaire about \name{}'s visual interface, and a semi-structured interview. 
The whole study lasted around 1 hour per participant.

\begin{table}[tb]
    \renewcommand{\arraystretch}{0.8}
    \small
    \centering
    \caption{Networks used for the controlled user study and case studies, where 
    $n$ and $l$ represent the numbers of nodes and links, respectively.}
    \label{table:network_info}
    \makeatletter
\def\thickhline{
  \noalign{\ifnum0=`}\fi\hrule \@height \thickarrayrulewidth \futurelet
   \reserved@a\@xthickhline}
\def\@xthickhline{\ifx\reserved@a\thickhline
               \vskip\doublerulesep
               \vskip-\thickarrayrulewidth
             \fi
      \ifnum0=`{\fi}}
\makeatother

\newlength{\thickarrayrulewidth}
\setlength{\thickarrayrulewidth}{2\arrayrulewidth}
\renewcommand{\tabcolsep}{4pt}

\begin{tabular}{llll}
\thickhline
 & \textit{Target Network $\GraphT$} & \textit{Background Network $\GraphB$} & \textit{\# of RFOs*} \\
\hline
\rowcolor[gray]{.95} Tutorial 1 & Price1 & Random1 & 0 \\
\rowcolor[gray]{.95} & ($n=100$, $l=294$) & ($n=100$, $l=471$) & \\
\rowcolor[gray]{1.0} Tutorial 2 & Random1 & Random2 & N/A \\
\rowcolor[gray]{1.0} & ($n=100$ , $l=471$) & ($n=100$ , $l=525$) & \\
\rowcolor[gray]{.95} Tutorial 3, & LC-multiple~\cite{yu2008high} & Combined-AP/MS~\cite{yu2008high} & 1 \\
\rowcolor[gray]{.95} Case Study 2 & ($n=1,536$, $l=2,925$) & ($n=1,622$, $l=9,070$) & \\
\rowcolor[gray]{1.0} Training & School-Day2~\cite{stehle2011high} & School-Day1~\cite{stehle2011high} & 2 \\
\rowcolor[gray]{1.0} & ($n=238$, $l=5,539$) & ($n=236$, $l=5,899$) & \\
\rowcolor[gray]{.95} Task A & Brain-Low~\cite{khambhati2018subgraphs} & Brain-High~\cite{khambhati2018subgraphs} & N/A \\
\rowcolor[gray]{.95} & ($n=233$ , $l=2,627$) & ($n=246$ , $l=3,355$) & \\
\rowcolor[gray]{1.0} Task B, & p2p-Gnutella08~\cite{snapnets} & Price2 & 0 \\
\rowcolor[gray]{1.0} Case Study 1 & ($n=6,301$, $l=20,777$) & ($n=6,301$, $l=18,897$) & \\
\rowcolor[gray]{.95} Task C & Dolphin~\cite{lusseau2003bottlenose} & Karate~\cite{zachary1977information} & 1 \\
\rowcolor[gray]{.95} & ($n=62$ , $l=159$) & ($n=34$ , $l=78$) & \\
\thickhline
\multicolumn{4}{@{}l}{* \# of RFOs in a network feature that highly contributes to the uniqueness in $\GraphT$.} \\

\end{tabular}

\end{table}

\begin{figure}[tb]
	\centering
	\captionsetup{farskip=0pt}
    \includegraphics[width=1.0\linewidth]{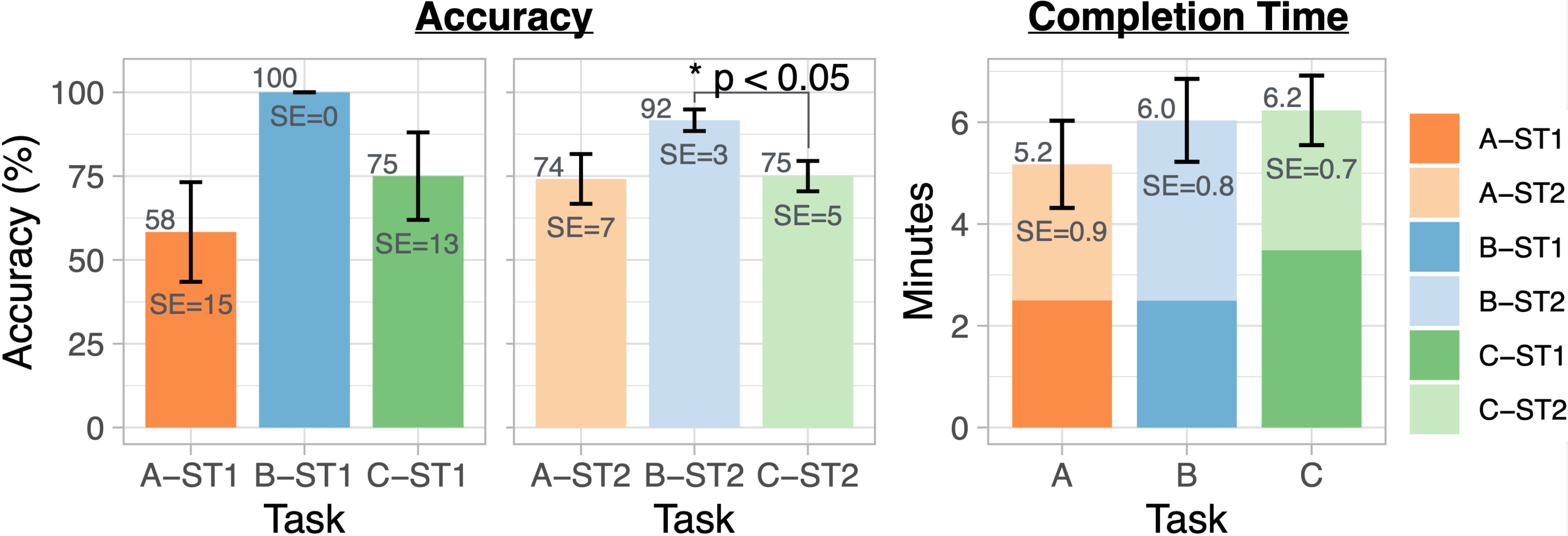}
    \caption{Accuracy (left) and completion time (right) for each subtask.}
	\label{fig:acc_time}
\end{figure}

\subsection{Results}
\label{sec:us_results}

This section reports our controlled study results including task accuracy, completion time, and participants' subjective feedback.

\textbf{Accuracy.}
The accuracy for each subtask is shown in \autoref{fig:acc_time}-left. 
Two network science experts independently rated participants' explanations in ST2 with a scale of 1 (the worst) to 5 (the best) based on correctness and comprehensiveness.
Weighted Cohen's kappa coefficient indicates high reliability of the ratings ($\kappa=0.83$, in the range of $0.81$--$1.00$: \textit{almost perfect agreement})~\cite{cohen1968weighted}. 

In general, Task B has the highest mean accuracy for both ST1 (100\%) and ST2 (92\%), which might be because the uniqueness of the target network can be understood easily with the base feature.
However, for ST1, a Cochran's Q test~\cite{mccrum2008correct} does not show any significant differences across tasks.
For ST2, a Friedman test~\cite{mccrum2008correct} reveals significant differences ($\chi^2 = 7.55$, $p < 0.05$).
A post-hoc analysis using Wilcoxon signed-rank exact test with Bonferroni correction~\cite{benavoli2016should} indicates that Task B has significantly higher accuracy than Task C ($p < 0.05$) that has the most difficulty.
Additionally, participants' scores of ST2 show a weak positive correlation (Pearson's correlation coefficient $\rho=0.31$) with the numbers of network centralities/measures they knew, which generally represent their level of expertise in network science. 
Thus, higher expertise seems to help provide better explanations.

\textbf{Completion Time.}
\autoref{fig:acc_time}-right shows the completion time for each task.
However, a Friedman test does not show any significant difference across tasks. 
There is a weak negative correlation ($\rho = -0.33$) between the completion times and the numbers of known network centralities/measures (i.e., the expertise helped finish tasks faster).
For Tasks A and B, ST2 (2.7 minutes and 3.5 minutes, respectively) took longer than ST1 (both 2.5 minutes). But for Task C, it is the opposite (ST1: 3.5 minutes, ST2: 2.7 minutes). 
From our observation, the reason might be that participants tried to find the explanation (ST2) before deciding their answer to ST1. 

\begin{figure}[tb]
	\centering
	\captionsetup{farskip=0pt}
    \includegraphics[width=1.0\linewidth]{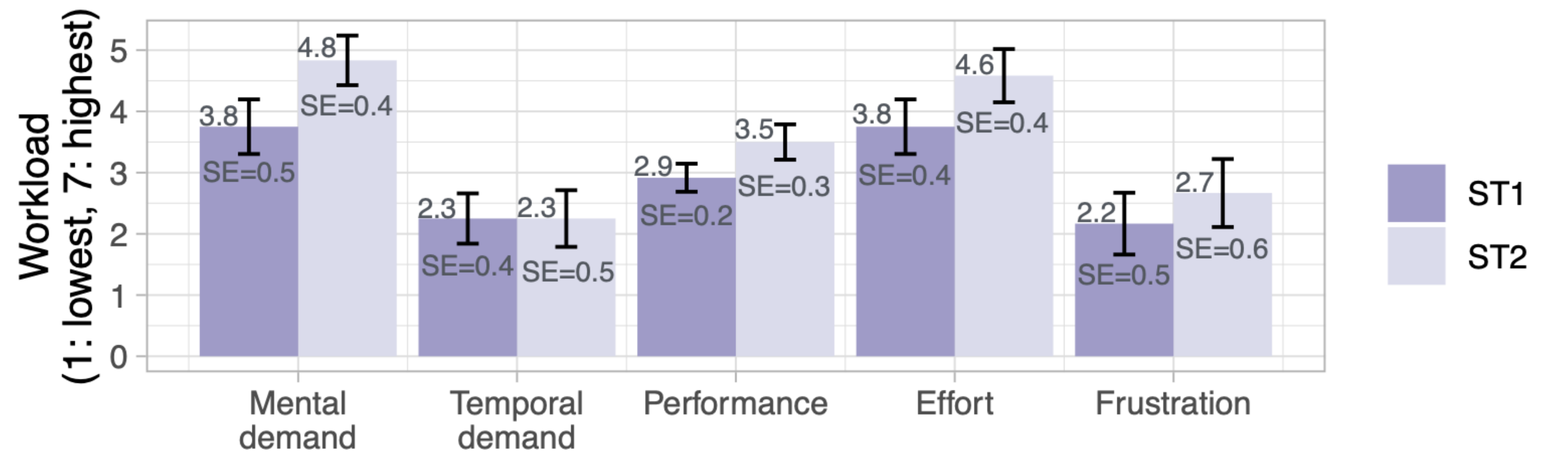}
    \caption{NASA TLX results (the lower the better).}
	\label{fig:nasa_tlx}
\end{figure}

\begin{figure}[tb]
	\centering
	\captionsetup{farskip=0pt}
    \includegraphics[width=1.0\linewidth]{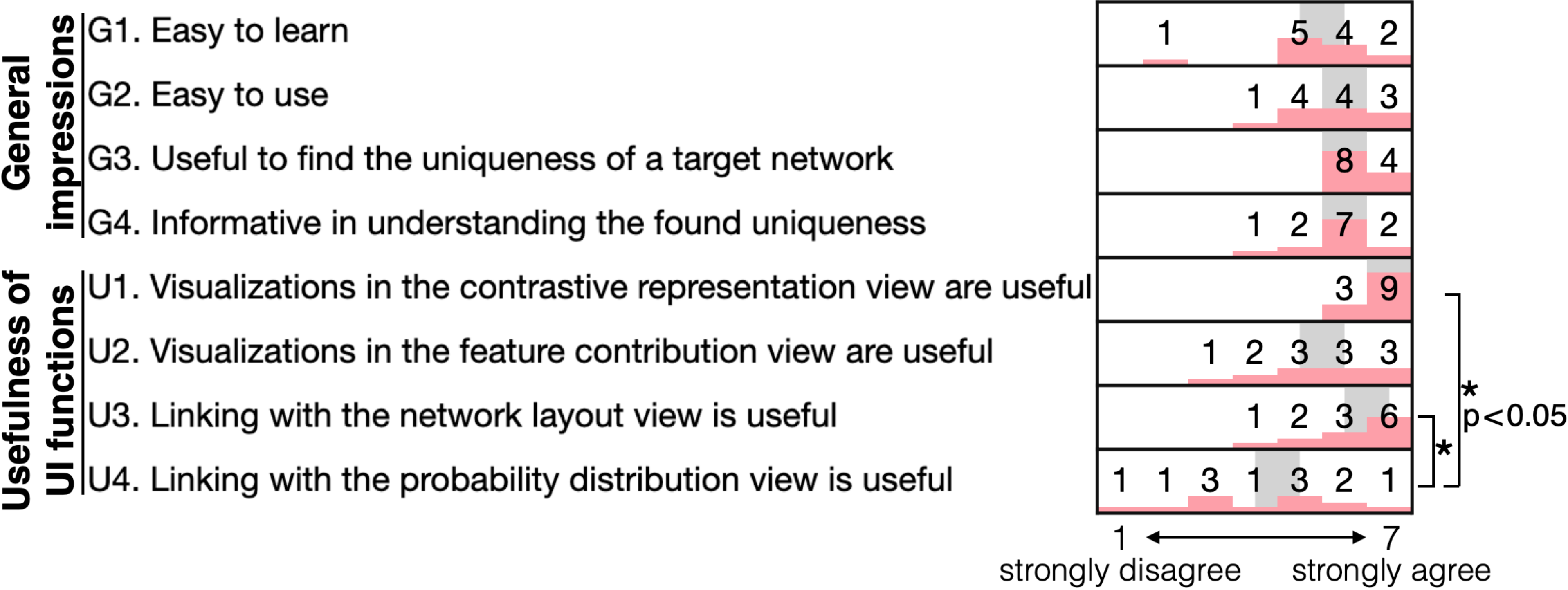}
    \caption{Histograms of participants' ratings on the overall impression and usefulness of each UI function (the higher the better). Numbers over the bins represent the frequency. Median ratings are indicated in gray.}
	\label{fig:survey_scores}
\end{figure}

\textbf{Subjective Feedback.}
\autoref{fig:nasa_tlx} lists participants' ratings with the NASA TLX. 
Generally, ST2 has higher mean values than ST1 in each task; however, a Wilcoxon signed-rank exact test does not show any significant difference in each pair of subtasks. 
Participants expressed relatively high mental demand and effort for performing the tasks, which is plausible because the network analysis needs high concentration.
\autoref{fig:survey_scores} shows the questionnaire results on the impression of \name{}. 
Overall, participants felt that \name{} is easy to learn, easy to use, and useful to perform ST1 and ST2. 
For the usefulness of each UI function, the contrastive representation, feature contribution, and network layout views receive high ratings, especially the contrastive representation view, whereas the probability distribution view has relatively low scores.
Also, a Friedman test ($\chi^2 = 18.0$, $p<0.001$) and a post-hoc analysis using the Wilcoxon signed-rank exact test with Bonferroni correction~\cite{benavoli2016should} show significant differences of the probability distribution view from the contrastive representation ($p<0.05$) and network layout ($p<0.05$) views on participants' ratings. 
One reason we obtained from the interviews is that the uniqueness can be identified and explained with other views, while the probability distribution view is not necessary, although it is helpful to confirm uniqueness.

During the interview, we collected the participants' preference for the feature representations in DeepGL (\autoref{fig:feat_repr}-a) and \name{} (\autoref{fig:feat_repr}-b). 
Ten out of 12 participants preferred \name{}'s representation because it is \q{visually more clear} (\p{4, 6}) and \q{more intuitive to understand} (other 8 participants). 
The rest of the participants preferred DeepGL's notation because using mathematical symbols has less ambiguity.
Eleven participants applauded the usefulness of \name{}'s visualization of intermediate computational results (\autoref{fig:step_by_step}), which was used to understand complicated network features: \q{Those are particularly useful because you can see the levels of how these [i.e., features] are getting computed like that} (\p{3}). 
Two participants with expert knowledge mentioned that they wanted to use the opposite order from the current representation (i.e., from left to right, placing RFOs first and then a base feature) because they mentally converted each network feature in this order.
However, others stated that they were used to understand each feature from a base feature. 

\section{Discussion}

We have presented \name{} and validated it with case studies and a controlled user study. Here, we provide a thematic discussion on additional aspects of \name{} as well as the studies. 

\textbf{Limitations in Visual Scalability.}
While the studies indicate the usefulness and effectiveness of \name{}, it is not without limitations. 
\name{} employs scatterplots, node-link diagrams, and heatmaps in its interface, but these techniques suffer from scalability issues. 
We can enhance these techniques with filtering, aggregation, and focus+context methods to mitigate the issues~\cite{cui2019visual}. 
A specific scalability issue in \name{} is the visual representation of network features, where we use rectangles in the feature contribution view, ellipses, lines with text labels.
This may limit the number of RFOs to display in a network feature.
However, this is not a major issue, because NRL methods (including DeepGL) that generate features based on relationships of node neighbors are generally utilized with only a few hops (typically 2 or 3) of neighbors~\cite{kipf2016semi,rossi2018deep}. 
Another issue could be caused when NRL produces a large number of features (e.g., 100 features). 
This issue can be addressed by only displaying features that highly contribute to cPCs as such features are most important to interpret the cNRL results.

\textbf{Ambiguity of Uniqueness.}
In spite of high mental demand and effort, participants achieved high accuracy in identifying (\textbf{Q1}) and explaining (\textbf{Q2}) the uniqueness in a target network when it actually exists (Tasks B and C). 
However, when a target network did not have clear uniqueness (Task A), the accuracy for ST1 was relatively low, though there was no significant difference (\autoref{sec:us_results}). 
Potential reasons might be associated with the ambiguity of uniqueness and participants' expectation, as noted by \p{1}: \q{It wasn't too difficult [to learn and use the contrastive representation view] but I had a question of how much separation is enough to define uniqueness.} 
While the contrastive representation view in Task A showed the similar scatteredness between target and background networks, participants were able to find some small regions that seemed to relate to the uniqueness if they had an expectation for uniqueness.
We found that all 5 participants who answered \textit{Yes} for Task A-ST1 did not provide a convincing explanation, with mean accuracy for Task A-ST2 $52\%$ (\autoref{fig:acc_time}). 
How to better define and inform a threshold of containing the uniqueness should be addressed in the further work. 

\textbf{Importance of Interpretability and cNRL.}
One notable result is that participants spent similar time in completing ST1 and ST2. 
This surprises us because we expected that ST1 would be finished much faster because they only needed to review the contrastive representation view and select an answer, while ST2 required the use of multiple views and writing an explanation. 
From our observation, we noticed that although they quickly recognized the uniqueness from the contrastive representation view, before selecting the answer, they tried to understand the reasons behind to convince themselves. 
This points out the importance of providing the interpretability in algorithms, including NRL and CL methods.  
This fact also influenced the mean accuracy of Task C-ST1.
Three participants chose \textit{I'm not sure} because they were not able to completely understand why the target network was unique while the potential uniqueness was found, which may be due to their lower expertise in network science. 

All the views except for the probability distribution view seemed to be useful according to participants.
From our interviews, several participants mentioned that for easier tasks (e.g., Task B) it was not necessary to use the probability distribution view; for more difficult tasks (e.g., Task C), the probability distribution view was not helpful to reveal the uniqueness.
This indicates the limitation of network comparison based on probability distributions, which is a popular analysis approach, and the necessity of more advanced embedding based approaches, such as cNRL.
Further, when asked about how to perform similar tasks without \name{}, participants provided approaches of either comparing probability distributions of basic network centralities or comparing laid-out networks. 
Also, they mentioned that they might be able to find the uniqueness with their stated approach but it would \q{be awful} (\p{9}) and \q{take longer} (\p{1}), and \q{I might miss some uniqueness} (\p{5}). In contrast, using \name{} is \q{much easier because it supports a lot of stuff you need to deal with... Comparing the target and background in the contrastive representation view is really helpful. If you see spreading patterns [of a target], it might be unique.} (\p{11}). 

\textbf{Usage with Other Algorithms.}
\name{} employs i-cNRL because of its interpretability; however, most of \name{} functionalities are generic enough to be well adapted with other NRL and CL methods in the architecture. 
For example, if the interpretability is not required, DeepGL can be replaced with GraphSAGE~\cite{hamilton2017inductive}, where only network features are changed. Thus, \name{} is still applicable by updating the visual representations of features in the feature contribution view.
Similarly, we can switch cPCA with the other CL methods, such as cVAE~\cite{abid2019contrastive,severson2019unsupervised} which can find uniqueness in a target dataset even when its data points and latent features have nonlinear relationships. 
As we cannot obtain the features' contributions, in this case, we can simply remove the heatmap from the feature contribution view. 
Also, once other interpretable CL methods are developed, we do not need any changes to integrate them into \name{}.
Another potential extension is cooperating with link feature learning, which is also supported by DeepGL. 
In this case, we just need to add visual encodings of links to the views of \name{}.

\textbf{Adaption for Application Domains.}
As presented, the linking with laid-out networks is important to intuitively understand uniqueness. 
Networks are often visualized in a specific manner according to the application domain.
For example, when analyzing brain networks, neuroscientists often use adjacency-matrix based visualizations or 2D/3D node-link diagrams~\cite{fornito2016fundamentals}.
This is because the former is useful to find correlated brain regions with matrix-reordering algorithms~\cite{behrisch2016matrix} and the latter can help relate analysis results to the actual locations in a brain. 
By customizing the network layout views, \name{} can support such analysis tasks in this specific domain. 
Also, as shown in \autoref{sec:us_results}, the way to understand network features generated by DeepGL is different by the analyst. 
Therefore, we should consider adding settings to customize the representation of network features based on the analyst's preference.

\section{Conclusion and Future Work}

We have presented \name{}, a visual analytics framework for network comparison, which utilizes two machine learning schemes---network representation learning and contrastive learning---together with an intuitive visual interface. 
\name{} provides the capability for effectively identifying and understanding unique characteristics of one network relative to another, supporting four key capabilities as outlined by \textbf{DIIF}. 
As our case studies indicate, \name{} promises to extract insights from networks found in various application domains. 
Our controlled user study also reflects the usefulness and effectiveness of \name{} with carefully-designed analysis tasks.  
In the future, to provide more variations for contrastive network analysis, we plan to extend the framework for comparison of two groups of multiple networks, including dynamic network comparison. 
We also wish to adapt \name{} to other cNRL algorithms (e.g., based on GraphSAGE~\cite{hamilton2017inductive}, cVAE~\cite{severson2019unsupervised}, etc.) and application domains (e.g., brain network analysis).

\acknowledgments{This research was partially carried out at FXPAL. This research is sponsored in part by the U.S. National Science Foundation through grant IIS-1741536, the Natural Sciences and Engineering Research Council of Canada through the Discovery Grant, and FXPAL through its internship program.}

\clearpage

\bibliographystyle{abbrv-doi}
\bibliography{00_main}

\begin{thebibliography}{10}

\bibitem{supp}
The supplementary materials: Videos, datasets, and user study materials.
\newblock \url{https://takanori-fujiwara.github.io/s/contrana/}.

\bibitem{abid2018exploring}
A.~Abid, M.~J. Zhang, V.~K. Bagaria, and J.~Zou.
\newblock Exploring patterns enriched in a dataset with contrastive principal
  component analysis.
\newblock {\em Nature Communications}, 9(1):2134, 2018.

\bibitem{abid2019contrastive}
A.~Abid and J.~Zou.
\newblock Contrastive variational autoencoder enhances salient features.
\newblock {\em arXiv:1902.04601}, 2019.

\bibitem{alper2013weighted}
B.~Alper, B.~Bach, N.~Henry~Riche, T.~Isenberg, and J.-D. Fekete.
\newblock Weighted graph comparison techniques for brain connectivity analysis.
\newblock In {\em Proc. CHI}, pp. 483--492, 2013.

\bibitem{aminikhanghahi2017survey}
S.~Aminikhanghahi and D.~J. Cook.
\newblock A survey of methods for time series change point detection.
\newblock {\em Knowledge and Information Systems}, 51(2):339--367, 2017.

\bibitem{andrews2009visual}
K.~Andrews, M.~Wohlfahrt, and G.~Wurzinger.
\newblock Visual graph comparison.
\newblock In {\em Proc. IV}, pp. 62--67, 2009.

\bibitem{bach2015small}
B.~Bach, N.~Henry-Riche, T.~Dwyer, T.~Madhyastha, J.-D. Fekete, and
  T.~Grabowski.
\newblock {Small MultiPiles}: Piling time to explore temporal patterns in
  dynamic networks.
\newblock {\em Computer Graphics Forum}, 34(3):31--40, 2015.

\bibitem{bach2013graphdiaries}
B.~Bach, E.~Pietriga, and J.-D. Fekete.
\newblock {GraphDiaries}: Animated transitions andtemporal navigation for
  dynamic networks.
\newblock {\em IEEE Trans. on Visualization and Computer Graphics},
  20(5):740--754, 2013.

\bibitem{bach2016time}
B.~Bach, C.~Shi, N.~Heulot, T.~Madhyastha, T.~Grabowski, and P.~Dragicevic.
\newblock {Time Curves}: Folding time to visualize patterns of temporal
  evolution in data.
\newblock {\em IEEE Trans. on Visualization and Computer Graphics},
  22(1):559--568, 2016.

\bibitem{barabasi2016network}
A.-L. Barab{\'a}si et~al.
\newblock {\em Network Science}.
\newblock Cambridge university press, 2016.

\bibitem{beck2017taxonomy}
F.~Beck, M.~Burch, S.~Diehl, and D.~Weiskopf.
\newblock A taxonomy and survey of dynamic graph visualization.
\newblock {\em Computer Graphics Forum}, 36(1):133--159, 2017.

\bibitem{behrisch2016matrix}
M.~Behrisch, B.~Bach, N.~Henry~Riche, T.~Schreck, and J.-D. Fekete.
\newblock Matrix reordering methods for table and network visualization.
\newblock {\em Computer Graphics Forum}, 35(3):693--716, 2016.

\bibitem{benavoli2016should}
A.~Benavoli, G.~Corani, and F.~Mangili.
\newblock Should we really use post-hoc tests based on mean-ranks?
\newblock {\em J. of Machine Learning Research}, 17(1):152--161, 2016.

\bibitem{bezerianos2010graphdice}
A.~Bezerianos, F.~Chevalier, P.~Dragicevic, N.~Elmqvist, and J.-D. Fekete.
\newblock {GraphDice}: A system for exploring multivariate social networks.
\newblock {\em Computer Graphics Forum}, 29(3):863--872, 2010.

\bibitem{bhanot2005optimizing}
G.~Bhanot, A.~Gara, P.~Heidelberger, E.~Lawless, et~al.
\newblock Optimizing task layout on the {Blue Gene/L} supercomputer.
\newblock {\em IBM J. of Research and Development}, 49(2.3):489--500, 2005.

\bibitem{bostock2011d3}
M.~Bostock, V.~Ogievetsky, and J.~Heer.
\newblock {D$^3$} data-driven documents.
\newblock {\em IEEE Trans. on Visualization and Computer Graphics},
  17(12):2301--2309, 2011.

\bibitem{cai2018comprehensive}
H.~Cai, V.~W. Zheng, and K.~C.-C. Chang.
\newblock A comprehensive survey of graph embedding: Problems, techniques, and
  applications.
\newblock {\em IEEE Trans. on Knowledge and Data Engineering},
  30(9):1616--1637, 2018.

\bibitem{chen2004content}
H.~Chen and B.~M. Sharp.
\newblock Content-rich biological network constructed by mining pubmed
  abstracts.
\newblock {\em BMC Bioinformatics}, 5(1):147, 2004.

\bibitem{coelho2020peckvis}
D.~Coelho, I.~Chase, and K.~Mueller.
\newblock {PeckVis}: A visual analytics tool to analyze dominance hierarchies
  in small groups.
\newblock {\em IEEE Trans. on Visualization and Computer Graphics},
  26(4):1650--1660, 2020.

\bibitem{cohen1968weighted}
J.~Cohen.
\newblock Weighted kappa: Nominal scale agreement provision for scaled
  disagreement or partial credit.
\newblock {\em Psychological Bulletin}, 70(4):213, 1968.

\bibitem{collins2007toward}
S.~R. Collins, P.~Kemmeren, X.-C. Zhao, J.~F. Greenblatt, et~al.
\newblock Toward a comprehensive atlas of the physical interactome of
  {Saccharomyces} cerevisiae.
\newblock {\em Molecular \& Cellular Proteomics}, 6(3):439--450, 2007.

\bibitem{crnovrsanin2014visualization}
T.~Crnovrsanin, C.~Muelder, R.~Faris, D.~Felmlee, and K.-L. Ma.
\newblock Visualization techniques for categorical analysis of social networks
  with multiple edge sets.
\newblock {\em Social Networks}, 37:56--64, 2014.

\bibitem{cui2019visual}
W.~Cui.
\newblock Visual analytics: A comprehensive overview.
\newblock {\em IEEE Access}, 7:81555--81573, 2019.

\bibitem{dal2017wavelet}
A.~Dal~Col, P.~Valdivia, F.~Petronetto, F.~Dias, C.~T. Silva, and L.~G. Nonato.
\newblock Wavelet-based visual analysis of dynamic networks.
\newblock {\em IEEE Trans. on Visualization and Computer Graphics},
  24(8):2456--2469, 2017.

\bibitem{dang2016timearcs}
T.~N. Dang, N.~Pendar, and A.~G. Forbes.
\newblock {TimeArcs}: Visualizing fluctuations in dynamic networks.
\newblock {\em Computer Graphics Forum}, 35(3):61--69, 2016.

\bibitem{palettable}
M.~Davis.
\newblock {Palettable}.
\newblock \url{https://jiffyclub.github.io/palettable/}.
\newblock Accessed: 2020-3-31.

\bibitem{dirie2019contrastive}
A.-H. Dirie, A.~Abid, and J.~Zou.
\newblock Contrastive multivariate singular spectrum analysis.
\newblock In {\em Proc. Allerton Conference}, pp. 1122--1127, 2019.

\bibitem{emmert2016fifty}
F.~Emmert-Streib, M.~Dehmer, and Y.~Shi.
\newblock Fifty years of graph matching, network alignment and network
  comparison.
\newblock {\em Information Sciences}, 346:180--197, 2016.

\bibitem{federico2011visual}
P.~Federico, W.~Aigner, S.~Miksch, F.~Windhager, and L.~Zenk.
\newblock A visual analytics approach to dynamic social networks.
\newblock In {\em Proc. of i-KNOW}, pp. 1--8, 2011.

\bibitem{fornito2016fundamentals}
A.~Fornito, A.~Zalesky, and E.~Bullmore.
\newblock {\em Fundamentals of Brain Network Analysis}.
\newblock Academic Press, 2016.

\bibitem{freire2010manynets}
M.~Freire, C.~Plaisant, B.~Shneiderman, and J.~Golbeck.
\newblock Manynets: an interface for multiple network analysis and
  visualization.
\newblock In {\em Proc. CHI}, pp. 213--222, 2010.

\bibitem{fujiwara2017visual}
T.~Fujiwara, J.-K. Chou, A.~M. McCullough, C.~Ranganath, and K.-L. Ma.
\newblock A visual analytics system for brain functional connectivity
  comparison across individuals, groups, and time points.
\newblock In {\em Proc. PacificVis}, pp. 250--259, 2017.

\bibitem{fujiwara2019incremental}
T.~Fujiwara, J.-K. Chou, S.~Shilpika, P.~Xu, L.~Ren, and K.-L. Ma.
\newblock An incremental dimensionality reduction method for visualizing
  streaming multidimensional data.
\newblock {\em IEEE Trans. on Visualization and Computer Graphics},
  26(1):418--428, 2020.

\bibitem{fujiwara2020supporting}
T.~Fujiwara, O.-H. Kwon, and K.-L. Ma.
\newblock Supporting analysis of dimensionality reduction results with
  contrastive learning.
\newblock {\em IEEE Trans. on Visualization and Computer Graphics},
  26(1):45--55, 2020.

\bibitem{fujiwara2018visual}
T.~Fujiwara, J.~K. Li, M.~Mubarak, C.~Ross, C.~D. Carothers, R.~B. Ross, and
  K.-L. Ma.
\newblock A visual analytics system for optimizing the performance of
  large-scale networks in supercomputing systems.
\newblock {\em Visual Informatics}, 2(1):98--110, 2018.

\bibitem{icnrl}
T.~Fujiwara, J.~Zhao, F.~Chen, Y.~Yu, and K.-L. Ma.
\newblock Interpretable contrastive learning for networks.
\newblock {\em arXiv:2005.12419}, 2020.

\bibitem{gaiteri2016genetic}
C.~Gaiteri, S.~Mostafavi, C.~J. Honey, P.~L. De~Jager, and D.~A. Bennett.
\newblock Genetic variants in alzheimer disease-molecular and brain network
  approaches.
\newblock {\em Nature Reviews Neurology}, 12(7):413, 2016.

\bibitem{ge2016rich}
R.~Ge and J.~Zou.
\newblock Rich component analysis.
\newblock In {\em Proc. ICML}, pp. 1502--1510, 2016.

\bibitem{gehlenborg2010visualization}
N.~Gehlenborg, S.~I. O'donoghue, N.~S. Baliga, A.~Goesmann, M.~A. Hibbs,
  H.~Kitano, et~al.
\newblock Visualization of omics data for systems biology.
\newblock {\em Nature Methods}, 7:S56--S68, 2010.

\bibitem{gleicher2018considerations}
M.~Gleicher.
\newblock Considerations for visualizing comparison.
\newblock {\em IEEE Trans. on Visualization and Computer Graphics},
  24(1):413--423, 2018.

\bibitem{gleicher2011visual}
M.~Gleicher, D.~Albers, R.~Walker, I.~Jusufi, C.~D. Hansen, and J.~C. Roberts.
\newblock Visual comparison for information visualization.
\newblock {\em Information Visualization}, 10(4):289--309, 2011.

\bibitem{goldenberg2010survey}
A.~Goldenberg, A.~X. Zheng, S.~E. Fienberg, and E.~M. Airoldi.
\newblock A survey of statistical network models.
\newblock {\em Foundations and Trends{\textregistered} in Machine Learning},
  2(2):129--233, 2010.

\bibitem{gove2019gragnostics}
R.~Gove.
\newblock Gragnostics: Fast, interpretable features for comparing graphs.
\newblock In {\em Proc. IV}, pp. 201--209, 2019.

\bibitem{grover2016node2vec}
A.~Grover and J.~Leskovec.
\newblock {node2vec}: Scalable feature learning for networks.
\newblock In {\em Proc. KDD}, pp. 855--864, 2016.

\bibitem{hajij2018visual}
M.~Hajij, B.~Wang, C.~Scheidegger, and P.~Rosen.
\newblock Visual detection of structural changes in time-varying graphs using
  persistent homology.
\newblock In {\em Proc. PacificVis}, pp. 125--134, 2018.

\bibitem{hamilton2017inductive}
W.~Hamilton, Z.~Ying, and J.~Leskovec.
\newblock Inductive representation learning on large graphs.
\newblock In {\em Proc. NIPS}, pp. 1024--1034, 2017.

\bibitem{harrigan2012egonav}
M.~Harrigan, D.~Archambault, P.~Cunningham, and N.~Hurley.
\newblock {EgoNav}: Exploring networks through egocentric spatializations.
\newblock In {\em Proc. AVI}, pp. 563--570, 2012.

\bibitem{harrower2003colorbrewer}
M.~Harrower and C.~A. Brewer.
\newblock Colorbrewer.org: An online tool for selecting colour schemes for
  maps.
\newblock {\em Cartographic Journal}, 40(1):27--37, 2003.

\bibitem{Hart1988development}
S.~G. Hart and L.~E. Staveland.
\newblock Development of nasa-tlx (task load index): Results of empirical and
  theoretical research.
\newblock In P.~A. Hancock and N.~Meshkati, eds., {\em Human Mental Workload},
  vol.~52 of {\em Advances in Psychology}, pp. 139--183. North-Holland, 1988.

\bibitem{henry2007nodetrix}
N.~Henry, J.-D. Fekete, and M.~J. McGuffin.
\newblock {NodeTrix}: a hybrid visualization of social networks.
\newblock {\em IEEE Trans. on Visualization and Computer Graphics},
  13(6):1302--1309, 2007.

\bibitem{hu2005efficient}
Y.~Hu.
\newblock Efficient, high-quality force-directed graph drawing.
\newblock {\em Mathematica Journal}, 10(1):37--71, 2005.

\bibitem{JiaNZ09}
Y.~{Jia}, F.~{Nie}, and C.~{Zhang}.
\newblock \href{10.1109/TNN.2009.2015760}{Trace Ratio Problem Revisited}.
\newblock {\em IEEE Trans. on Neural Networks}, 20(4):729--735, 2009.

\bibitem{john2019visual}
M.~John and M.~Baumann.
\newblock A visual approach for the comparative analysis of character networks
  in narrative texts.
\newblock In {\em Proc. PacificVis}, pp. 247--256, 2019.

\bibitem{jolliffe1986principal}
I.~T. Jolliffe.
\newblock Principal component analysis and factor analysis.
\newblock In {\em Principal Component Analysis}, pp. 115--128. Springer, 1986.

\bibitem{kerpedjiev2018higlass}
P.~Kerpedjiev, N.~Abdennur, F.~Lekschas, C.~McCallum, K.~Dinkla, H.~Strobelt,
  J.~M. Luber, S.~B. Ouellette, A.~Azhir, N.~Kumar, et~al.
\newblock {HiGlass}: web-based visual exploration and analysis of genome
  interaction maps.
\newblock {\em Genome biology}, 19(1):1--12, 2018.

\bibitem{kesavan2020visual}
S.~P. Kesavan, T.~Fujiwara, J.~K. Li, C.~Ross, M.~Mubarak, C.~D. Carothers,
  R.~B. Ross, and K.-L. Ma.
\newblock A visual analytics framework for reviewing streaming performance
  data.
\newblock In {\em Proc. PacificVis}, pp. 206--2015, 2020.

\bibitem{khambhati2018subgraphs}
A.~N. Khambhati, J.~D. Medaglia, E.~A. Karuza, S.~L. Thompson-Schill, and D.~S.
  Bassett.
\newblock Subgraphs of functional brain networks identify dynamical constraints
  of cognitive control.
\newblock {\em PLOS Computational Biology}, 14(7):e1006234, 2018.

\bibitem{kipf2016semi}
T.~N. Kipf and M.~Welling.
\newblock Semi-supervised classification with graph convolutional networks.
\newblock {\em arXiv:1609.02907}, 2016.

\bibitem{koop2013visual}
D.~Koop, J.~Freire, and C.~T. Silva.
\newblock Visual summaries for graph collections.
\newblock In {\em Proc. PacificVis}, pp. 57--64, 2013.

\bibitem{kwon2016axisketcher}
B.~C. Kwon, H.~Kim, E.~Wall, J.~Choo, H.~Park, and A.~Endert.
\newblock {AxiSketcher}: Interactive nonlinear axis mapping of visualizations
  through user drawings.
\newblock {\em IEEE Trans. on Visualization and Computer Graphics},
  23(1):221--230, 2017.

\bibitem{kwon2017would}
O.-H. Kwon, T.~Crnovrsanin, and K.-L. Ma.
\newblock What would a graph look like in this layout? a machine learning
  approach to large graph visualization.
\newblock {\em IEEE Trans. on Visualization and Computer Graphics},
  24(1):478--488, 2018.

\bibitem{lariviere2006canadian}
V.~Larivi{\`e}re, Y.~Gingras, and {\'E}.~Archambault.
\newblock Canadian collaboration networks: A comparative analysis of the
  natural sciences, social sciences and the humanities.
\newblock {\em Scientometrics}, 68(3):519--533, 2006.

\bibitem{lee2019dynamic}
A.~Lee, D.~Archambault, and M.~Nacenta.
\newblock {Dynamic Network Plaid}: A tool for the analysis of dynamic networks.
\newblock In {\em Proc. of ACM CHI}, pp. 1--14, 2019.

\bibitem{snapnets}
J.~Leskovec and A.~Krevl.
\newblock {SNAP Datasets}: {Stanford} large network dataset collection.
\newblock \url{http://snap.stanford.edu/data}, 2014.
\newblock Accessed: 2020-4-10.

\bibitem{liu2017egocomp}
D.~Liu, F.~Guo, B.~Deng, H.~Qu, and Y.~Wu.
\newblock {egoComp}: A node-link-based technique for visual comparison of
  ego-networks.
\newblock {\em Information Visualization}, 16(3):179--189, 2017.

\bibitem{liu2009efficient}
L.~Liu, J.~Xu, D.~Russell, P.~Townend, and D.~Webster.
\newblock Efficient and scalable search on scale-free p2p networks.
\newblock {\em Peer-to-Peer Networking and Applications}, 2(2):98--108, 2009.

\bibitem{liu2014topicpanorama}
S.~Liu, X.~Wang, J.~Chen, J.~Zhu, and B.~Guo.
\newblock {TopicPanorama}: A full picture of relevant topics.
\newblock In {\em Proc. VAST}, pp. 183--192, 2014.

\bibitem{liu2015effects}
X.~Liu and H.-W. Shen.
\newblock The effects of representation and juxtaposition on graphical
  perception of matrix visualization.
\newblock In {\em Proc. CHI}, pp. 269--278, 2015.

\bibitem{lusseau2003bottlenose}
D.~Lusseau, K.~Schneider, O.~J. Boisseau, P.~Haase, et~al.
\newblock The bottlenose dolphin community of doubtful sound features a large
  proportion of long-lasting associations.
\newblock {\em Behavioral Ecology and Sociobiology}, 54(4):396--405, 2003.

\bibitem{mccrum2008correct}
E.~McCrum-Gardner.
\newblock Which is the correct statistical test to use?
\newblock {\em British J. of Oral and Maxillofacial Surgery}, 46(1):38--41,
  2008.

\bibitem{moshiri2018dual}
N.~Moshiri.
\newblock The dual-{Barab{\'a}si-Albert} model.
\newblock {\em arXiv:1810.10538}, 2018.

\bibitem{murugesan2019state}
S.~Murugesan, K.~Bouchard, J.~Brown, M.~Kiran, D.~Lurie, B.~Hamann, and G.~H.
  Weber.
\newblock State-based network similarity visualization.
\newblock {\em Information Visualization}, 19(2):96--113, 2020.

\bibitem{newman2018networks}
M.~Newman.
\newblock {\em Networks}.
\newblock Oxford university press, 2018.

\bibitem{prvzulj2007biological}
N.~Pr{\v{z}}ulj.
\newblock Biological network comparison using graphlet degree distribution.
\newblock {\em Bioinformatics}, 23(2):e177--e183, 2007.

\bibitem{reguly2006comprehensive}
T.~Reguly, A.~Breitkreutz, L.~Boucher, B.-J. Breitkreutz, et~al.
\newblock Comprehensive curation and analysis of global interaction networks in
  saccharomyces cerevisiae.
\newblock {\em J. of Biology}, 5(4):11, 2006.

\bibitem{richman1986rotation}
M.~B. Richman.
\newblock Rotation of principal components.
\newblock {\em J. of Climatology}, 6(3):293--335, 1986.

\bibitem{rossi2018deep}
R.~A. Rossi, R.~Zhou, and N.~Ahmed.
\newblock Deep inductive graph representation learning.
\newblock {\em IEEE Trans. on Knowledge and Data Engineering}, 32(3):438--452,
  2020.

\bibitem{rufiange2013diffani}
S.~Rufiange and M.~J. McGuffin.
\newblock {DiffAni}: Visualizing dynamic graphs with a hybrid of difference
  maps and animation.
\newblock {\em IEEE Trans. on Visualization and Computer Graphics},
  19(12):2556--2565, 2013.

\bibitem{sacha2016visual}
D.~Sacha, L.~Zhang, M.~Sedlmair, J.~A. Lee, J.~Peltonen, D.~Weiskopf, S.~C.
  North, and D.~A. Keim.
\newblock Visual interaction with dimensionality reduction: A structured
  literature analysis.
\newblock {\em IEEE Trans. on Visualization and Computer Graphics},
  23(1):241--250, 2017.

\bibitem{severson2019unsupervised}
K.~A. Severson, S.~Ghosh, and K.~Ng.
\newblock Unsupervised learning with contrastive latent variable models.
\newblock In {\em Proc. AAAI}, vol.~33, pp. 4862--4869, 2019.

\bibitem{shi2015brainquest}
L.~Shi, H.~Tong, and X.~Mu.
\newblock Brainquest: Perception-guided brain network comparison.
\newblock In {\em Proc. ICDM}, pp. 379--388, 2015.

\bibitem{mplcolormaps}
N.~Smith and S.~van~der Walt.
\newblock {mpl colormaps}.
\newblock \url{https://bids.github.io/colormap/}.
\newblock Accessed: 2020-3-31.

\bibitem{stehle2011high}
J.~Stehl{\'e}, N.~Voirin, A.~Barrat, C.~Cattuto, et~al.
\newblock High-resolution measurements of face-to-face contact patterns in a
  primary school.
\newblock {\em PlOS ONE}, 6(8), 2011.

\bibitem{tam2016analysis}
G.~K. Tam, V.~Kothari, and M.~Chen.
\newblock An analysis of machine-and human-analytics in classification.
\newblock {\em IEEE Trans. on Visualization and Computer Graphics},
  23(1):71--80, 2017.

\bibitem{tantardini2019comparing}
M.~Tantardini, F.~Ieva, L.~Tajoli, and C.~Piccardi.
\newblock Comparing methods for comparing networks.
\newblock {\em Scientific Reports}, 9(1):17557, 2019.

\bibitem{torgerson1952}
W.~S. Torgerson.
\newblock Multidimensional scaling: I. {T}heory and method.
\newblock {\em Psychometrika}, 17(4):401--419, 1952.

\bibitem{turkay2017designing}
C.~Turkay, E.~Kaya, S.~Balcisoy, and H.~Hauser.
\newblock Designing progressive and interactive analytics processes for
  high-dimensional data analysis.
\newblock {\em IEEE Trans. on Visualization and Computer Graphics},
  23(1):131--140, 2017.

\bibitem{van2016reducing}
S.~van~den Elzen, D.~Holten, J.~Blaas, and J.~J. van Wijk.
\newblock Reducing snapshots to points: A visual analytics approach to dynamic
  network exploration.
\newblock {\em IEEE Trans. on Visualization and Computer Graphics},
  22(1):1--10, 2016.

\bibitem{von2009visual}
T.~von Landesberger, M.~Gorner, and T.~Schreck.
\newblock Visual analysis of graphs with multiple connected components.
\newblock In {\em Proc. VAST}, pp. 155--162, 2009.

\bibitem{wilkins2009hares}
M.~R. Wilkins.
\newblock Hares and tortoises: The high-versus low-throughput proteomic race.
\newblock {\em Electrophoresis}, 30(S1):S150--S155, 2009.

\bibitem{yang2017blockwise}
X.~Yang, L.~Shi, M.~Daianu, H.~Tong, Q.~Liu, and P.~Thompson.
\newblock Blockwise human brain network visual comparison using nodetrix
  representation.
\newblock {\em IEEE Trans. on Visualization and Computer Graphics},
  23(1):181--190, 2017.

\bibitem{yoghourdjian2018graph}
V.~Yoghourdjian, T.~Dwyer, K.~Klein, K.~Marriott, and M.~Wybrow.
\newblock {Graph Thumbnails}: Identifying and comparing multiple graphs at a
  glance.
\newblock {\em IEEE Trans. on Visualization and Computer Graphics},
  24(12):3081--3095, 2018.

\bibitem{yu2008high}
H.~Yu, P.~Braun, M.~A. Y{\i}ld{\i}r{\i}m, I.~Lemmens, et~al.
\newblock High-quality binary protein interaction map of the yeast interactome
  network.
\newblock {\em Science}, 322(5898):104--110, 2008.

\bibitem{zachary1977information}
W.~W. Zachary.
\newblock An information flow model for conflict and fission in small groups.
\newblock {\em J. of Anthropological Research}, 33(4):452--473, 1977.

\bibitem{zhang2018network}
D.~Zhang, J.~Yin, X.~Zhu, and C.~Zhang.
\newblock Network representation learning: A survey.
\newblock {\em IEEE Trans. on Big Data}, 2018.

\bibitem{zhang2018deep}
Z.~Zhang, P.~Cui, and W.~Zhu.
\newblock Deep learning on graphs: A survey.
\newblock {\em IEEE Trans. on Knowledge and Data Engineering}, 2020 (Early
  Access).

\bibitem{zhao2016egocentric}
J.~Zhao, M.~Glueck, F.~Chevalier, Y.~Wu, and A.~Khan.
\newblock Egocentric analysis of dynamic networks with egolines.
\newblock In {\em Proc. of ACM CHI}, pp. 5003--5014, 2016.

\bibitem{Zhao15}
J.~Zhao, Z.~Liu, M.~Dontcheva, A.~Hertzmann, and A.~Wilson.
\newblock {MatrixWave}: Visual comparison of event sequence data.
\newblock In {\em Proc. of ACM CHI}, p. 259–268, 2015.

\bibitem{zou2013contrastive}
J.~Y. Zou, D.~J. Hsu, D.~C. Parkes, and R.~P. Adams.
\newblock Contrastive learning using spectral methods.
\newblock In {\em Proc. NIPS}, pp. 2238--2246, 2013.

\end{thebibliography}
\end{document}